\documentclass[review,5p]{elsarticle}
\usepackage{hyperref}
\usepackage[utf8]{inputenc}
\usepackage{amsmath}
\usepackage{amssymb}
\usepackage{bm}
\usepackage{cleveref}

\usepackage{pdfpages}
\usepackage{graphicx}
\usepackage[american]{babel}
\usepackage[amssymb]{SIunits}
\usepackage{subfigure}
\usepackage{notoccite}
\usepackage{afterpage}
\biboptions{sort&compress}

\usepackage{tikz}
\usepackage{ulem}
\usetikzlibrary{calc}
\usetikzlibrary{plotmarks}
\usepackage{pgfplots}
\usepackage{etoolbox}
\patchcmd{\normalsize}{13.6}{13}{}{}
\pgfplotsset{
tick label style={font=\tiny},
label style={font=\small},
legend style={font=\tiny}
}
\pgfplotsset{compat=1.9}

\begin{document}

\begin{frontmatter}
 
\title{Understanding the mechanisms of electroplasticity from a crystal plasticity perspective}

\author[add]{Arka Lahiri\corref{correspondingauthor}}
\ead{a.lahiri@mpie.de, arkalahiri2009@gmail.com}

\author[add,add1]{Pratheek Shanthraj}
\ead{pratheek.shanthraj@manchester.ac.uk}

\author[add]{Franz Roters}%
\ead{f.roters@mpie.de}

\address[add]{%
Max-Planck-Institut f{\"u}r Eisenforschung GmBH, 
Max-Planck-Stra$\beta$e 1, 40237 D{\"u}sseldorf, Germany.
}%
\address[add1]{%
School of Materials, The University of Manchester, 
M13 9PL, UK
}%
\cortext[correspondingauthor]{Corresponding author}


\begin{abstract}
Electroplasticity is defined as the reduction in flow stress of 
a material undergoing deformation on passing an electrical pulse 
through it. The lowering of flow stress during electrical pulsing has been attributed to 
a combination of three mechanisms: softening due to Joule-heating of the material, 
de-pinning of dislocations from paramagnetic obstacles, and the electron-wind force acting on dislocations.  
However, there is no consensus in literature regarding the relative magnitudes 
of the reductions in flow stress resulting from each of these mechanisms. In this paper, we 
extend a dislocation density based crystal plasticity model to incorporate the 
mechanisms of electroplasticity and perform simulations where a single 
electrical pulse is applied during compressive deformation of a 
polycrystalline FCC material with random texture. 
We analyze the reductions in flow stress to understand the relative importance of the different mechanisms of electroplasticity
and delineate their dependencies on the various parameters 
related to electrical pulsing and dislocation motion.
Our study establishes that the reductions in flow stress are largely due to the 
mechanisms of de-pinning of dislocations from paramagnetic obstacles and Joule-heating,
with their relative dominance determined by the specific choice of crystal plasticity 
parameters corresponding to the particular material of interest. 

\end{abstract}

\end{frontmatter}

\section{Introduction}
Electroplasticity (henceforth called as ``EP'') is the phenomenon where a material undergoing deformation displays
a drop in flow stress whenever subjected to an electrical pulse.
The discovery of this phenomenon can be credited to Troitskii and Likhtman~\cite{Troitskii1963}
who first observed the reductions in flow stress while passing current pulses through Zn single crystals.
Since then it has been recognized that repeated application of the electrical pulses in quick succession during deformation 
lower the flow stress not only during the pulses but also between the pulses~\cite{Roth2008, Salandro2009, Salandro2010}.
Thus, repeated electropulsing during deformation mimics the attributes of hot working,
albeit at a much lower energy cost. 
This has prompted development of industrial manufacturing paradigms
like Electrically-assisted manufacturing (EAM) and Electroplastic manufacturing 
processing (EPMP) which leverage the phenomenon of EP on an industrial scale. 
There are several reviews of EAM, EPMP, and the phenomenon of EP which can serve as useful references 
in this regard~\cite{Sprecher1986, Guan2010, Nguyen2015, Ruszkiewicz2017}.

Even as there are increased efforts to harness the benefits of EP in the manufacturing industry,
the mechanisms contributing to EP continue to be poorly understood. Researchers
over the years have proposed several different theories to explain the reductions in flow stress during electropulsing
but without achieving any consensus  
about the mechanism which dominates the phenomenon of EP.
The earliest theory to explain the electroplastic softening is that of Joule-heating; the electrical energy passed 
to the material is converted to heat, leading to thermal softening of the material. This is the 
simplest explanation of EP but it is far from being unanimously accepted. 
Several experiments report the temperature rise due to a single electrical pulse to be too small to be commensurate 
with the reductions in flow stress~\cite{Sprecher1986, Conrad1990, Conrad2002}, while some later studies attribute the observed softening 
during electropulsing solely to Joule-heating~\cite{Goldman1981, Magargee2013}. 
The ambiguity surrounding Joule heating as the dominant 
mechanism for EP prompted development of theories which could explain the 
reductions in flow stress without invoking a rise in temperature of the material.
Conrad and co-workers~\cite{Sprecher1986, Conrad1990}, present the first among the athermal theories, which is based on 
the transfer of momentum from the flowing electrons to the dislocations. It is also 
known as the ``electron-wind force'' theory and was conjectured to be the 
principal contributor to EP
till Molotskii et al.,~\cite{Molotskii1995} presented an analysis which demonstrated its effect to be small compared to the reductions in flow stress
observed during experiments. Molotskii and co-workers~\cite{Molotskii1995,Molotskii2000} also present a different explanation for the reductions in flow stress.
They claim that the induced magnetic field due to the applied current alter the electronic states of the
bonds between the obstacles and the dislocation cores which promote de-pinning of dislocations from such obstacles.
Their theory requires the obstacles to be paramagnetic in character. The most prominent example of such 
obstacles are forest dislocations~\cite{Molotskii1995_2}, which constitute one of the biggest contributors to flow hardening. 
Molotskii et al.,~\cite{Molotskii1995, Molotskii2000} also present an analysis 
of the reductions in flow stress due to such an effect and find the
softening to be quite substantial compared to the two 
earlier mechanisms. 

To date, the theories of Joule-heating, electron-wind force, and de-pinning from paramagnetic obstacles 
are the most commonly invoked explanations for instances of EP observed in different materials.
The lack of agreement within the scientific community as to which mechanism dominates the electroplastic 
behaviour is partly because of the difficulty involved in experimentally validating the de-pinning of dislocations
and the electron-wind force on dislocations due to electropulsing. 
There are in-situ TEM studies where dislocation motion is observed 
during electropulsing of thin-films~\cite{Livesay1992,Vdovin1988}.    
But more recent studies~\cite{Kang2016,Kim2016}
claim that no difference in dislocation activity is observed 
under current pulsing. It has to be noted that in all these observations 
there is no concurrent applied strain while electropulsing. Thus, in these studies,
the imaged dislocations are static while the current pulses are applied and hence the tests
may fail to recognize whether larger segments of dislocations have been freed due to de-pinning. 
Similarly, the validity of the electron-wind force theory also cannot be ascertained with certainty through such experiments
as the transfer of momentum from the electrons to the dislocations may not constitute a high enough force by itself to cause dislocation motion.

With there being no clear understanding of the relative magnitudes
of softening induced by different mechanisms of EP
through experiments, modeling could play a key role in resolving this issue. 
There exists several crystal plasticity studies in literature which have
tried to model electrically assisted forming~\cite{Li2009,Roh2014,Hariharan2015, Krishnaswamy2017, Kim2018}.  
In a couple of such studies~\cite{Roh2014,Kim2018} constitutive models are 
presented which explain the envelope of the global stress-strain curve during frequent 
electropulsing without 
modeling the reductions in flow stress during electropulsing. The global softening of the material 
is captured phenomenologically without the consideration of any physical mechanisms other than Joule-heating.
On the other hand, in~\cite{Hariharan2015, Krishnaswamy2017}, the reductions in flow stress during pulsing 
are also modeled along with the global softening of the material due to repeated electropulsing.
These phenomenological crystal plasticity models only allow an empirical consideration of the
softening due to Joule-heating and electron-dislocation interactions and do not consider
any particular athermal mechanism of softening.
As these models do not implement the possible mechanisms of EP explicitly,
the questions regarding the relative importance of the proposed mechanisms
continue to remain unresolved.

In this paper, our objective is to understand 
the relative importance of the three theories 
of EP, namely, thermal softening, electron-wind force, 
and paramagnetic depinning of dislocations, in producing the 
reductions in flow stress during electropulsing. 
In order to achieve this,
we employ a dislocation density based crystal plasticity model and 
extend it to include the different mechanisms proposed for EP.
We then perform simulations of uniaxial compression 
of a representative polycrystalline sample where we pass a single electrical pulse
during the loading process and analyze the reduction in flow stress achieved when each 
of these mechanisms is active. 
This allows us to develop an understanding 
of the reductions in flow stress caused by each of these mechanisms and we probe
their dependencies on relevant parameters of the crystal plasticity model.
A major novelty of our approach from the
previous attempts at modeling EP is the fact that we 
use a dislocation density based crystal plasticity model which provides 
a physical framework for the introduction of the mechanisms of EP,
in contrast to the phenomenological and empirical approaches of earlier studies.
This also means that the crystal plasticity parameters employed in our study have a 
physical significance and can be determined from literature.
We utilize this flexibility to select a parameter set for our simulations 
which is representative of a generic FCC material and hence our conclusions are not limited 
to any one particular material displaying EP.
Our paper is organized 
in the following manner. First, we present the crystal 
plasticity model, which is followed by sections which extend the model to include athermal and 
thermal (Joule-heating) mechanisms of EP. We then present our results,
following which we discuss the implications of our study and lay out the possibilities for generalizing 
the model further to materials belonging to other crystal structures.

\section{Crystal plasticity model}\label{cpmodel}
 We will be using a dislocation density based crystal plasticity model which is described in detail by Wong et al.,~\cite{Wong2016}.
 As discussed in the previous section, all the mechanisms of EP are mediated exclusively by 
 dislocation motion. So, in our discussion, we are not going to invoke the other mechanisms which contribute to plasticity,
 e.g., twinning and transformation induced plasticity (TWIP, TRIP).  
 Our discussion will begin with a review of the kinematic and constitutive relationships of the model.
 
 \subsection{Kinematic and constitutive relationships}
 The imposed deformation gradient $\boldsymbol{F}$ is decomposed 
 into elastic ($\boldsymbol{F_e}$) and plastic ($\boldsymbol{F_p}$) contributions following \cite{Lee1969},
 \begin{align}
  \boldsymbol{F} =\boldsymbol{ F_e F_p}.
  \label{F_decomp}
 \end{align}
 The stress is computed from the elastic strain by assuming a linear elastic material,     
 \begin{align}
  \boldsymbol{S}= \mathbb{C} \left(\boldsymbol{F_e}^T \boldsymbol{F_e} - \boldsymbol{I} \right) / 2,
  \label{S_define}
 \end{align}
 where, $\boldsymbol{S}$ is the second Piola-Kirchhoff stress tensor and $\mathbb{C}$ is the 
 elastic tensor. The plastic velocity gradient ($\boldsymbol{L_p}$) for a single grain is determined by the stress tensor ($\boldsymbol{S}$)
 and the variables which define the microstructure ($\boldsymbol{\xi}$) as,
 \begin{align}
  \boldsymbol{L_p} = \sum_{\alpha=1}^{N_s} \dot{\gamma^\alpha}\left( \boldsymbol{S, \xi} \right) \boldsymbol{m^\alpha} \otimes \boldsymbol{n^\alpha},
  \label{lp_evol}
 \end{align}
 where $\boldsymbol{m^\alpha}$ denotes the slip direction and $\boldsymbol{n^\alpha}$ denotes the slip plane normal of the slip system $\alpha$.  
 $\dot{\gamma^\alpha}$ denote the shear rates on the individual slip systems denoted by $\alpha$. $N_s$ denotes the total
 number of slip systems. The exact form of the dependence
 of $\dot{\gamma^\alpha}$ on $\boldsymbol{S}$ and $\boldsymbol{\xi}$ will be delineated in the next subsection.
 As discussed earlier, Eq.~\ref{lp_evol} considers dislocation motion to be the sole mechanism responsible for plasticity.  
 $\boldsymbol{L_p}$ governs the evolution of the plastic deformation gradient through, 
 \begin{align}
 \dot{\boldsymbol{F_p}} = \boldsymbol{L_p F_p}, 
 \label{Fp_evol}
 \end{align}
 where, $\dot{\boldsymbol{F_p}}$ denotes the rate of change of the plastic deformation gradient with time. 
 The ~\cref{F_decomp,S_define,lp_evol,Fp_evol}, can be combined to write,
 \begin{align}
  \boldsymbol{P(x)} = \boldsymbol{F}\boldsymbol{S} = \boldsymbol{f\left(x, F, \xi \right)},
  \label{const_law}
 \end{align}
 where, $\boldsymbol{P}$ is the first Piola-Kirchhoff stress tensor determined from $\boldsymbol{S}$.
 Eq.~\ref{const_law} is the crux of the crystal plasticity model and is coupled to the balance of 
 linear momentum through,
 \begin{align}
  \nabla \cdot \boldsymbol{P} = 0,
  \label{lin_mom}
 \end{align}
 to simulate a Representative Volume Element (RVE) under static equilibrium. The numerical schemes employed to 
 solve these equations are described in detail elsewhere~\cite{Diehl2016}.   
  
 \subsection{Microstructure ($\xi$) and shear rates $\dot{\gamma^\alpha}$}
 The microstructure ($\boldsymbol{\xi}$) of the material is described by the mobile edge dislocation density
 denoted by $\rho_m$ and the immobile dipole dislocation density $\rho_d$.
 The motion of the mobile dislocations determines the shear rate $\dot{\gamma^\alpha}$ on the slip system 
 $\alpha$, as given by the Orowan equation,
 \begin{align}
  \dot{\gamma^\alpha} = \rho_m^\alpha b_s v_0 \exp\left[ - \dfrac{Q_s}{k_\text{B} T} \left\{ 1 - \left( \dfrac{\tau_{eff}^\alpha}{\tau_{sol}}\right)^p \right\}^q \right] \textrm{sign}(\tau^\alpha),
  \label{Orowan_eq}
 \end{align}
 which assumes dislocation glide to be controlled by thermal activation.
 In Eq.~\ref{Orowan_eq}, $\rho_m^\alpha$ denotes the mobile 
 dislocation density in the slip system $\alpha$,
 $Q_s$ is the activation energy for slip, 
 $b_s$ is the magnitude of the Burgers vector, 
 $k_\text{B}$ represents the Boltzmann constant and $T$ is the temperature,  
 $\tau_{sol}$ is the stress required to overcome short range obstacles at $0K$,
 $v_0$ is the dislocation glide velocity prefactor.
 $\tau_{eff}^\alpha$ is the effective resolved shear stress written as, 
 $\tau_{eff}^\alpha = |\tau^\alpha| - \tau_{pass}^\alpha$,
 when $|\tau^\alpha| > \tau_{pass}^\alpha$, and $\tau_{eff}^\alpha = 0$, otherwise. 
 $\tau^\alpha$ is the resolved applied shear stress on the slip system $\alpha$
 given by $\tau^\alpha = {\boldsymbol{F_e}}^T \boldsymbol{F_e} \boldsymbol{S} \left( {\boldsymbol{m}}^\alpha \otimes {\boldsymbol{n}}^\alpha \right)$
 and $\tau_{pass}^\alpha$ is the passing stress experienced by the mobile dislocations in 
 the slip system $\alpha$, due to the long range elastic strain fields of the dislocations,
 defined as,
 \begin{align}
  \tau_{pass}^\alpha = G b_s \left[ \sum_{{\alpha'}=1}^{N_s} \xi_{\alpha \alpha'} \left( \rho_e^{\alpha'} + \rho_d^{\alpha'} \right)  \right]^{1/2},
  \label{tau_pass_def}
 \end{align}
  where, $G$ is the shear modulus, and $\xi_{\alpha \alpha'}$  is the interaction matrix of the slip systems.
  It is important to note that the first instance when $|\tau^\alpha|$ 
  is greater than $\tau_{pass}^\alpha$ 
  defines yielding on the slip system $\alpha$, where the corresponding value of 
  $\tau_{pass}^\alpha$ is the slip system level yield stress.
  $\rho_m^{\alpha'}$ and $\rho_d^{\alpha'}$
  are the mobile edge and immobile dipole dislocation densities,
  respectively, whose evolutions are governed by,
  \begin{align}
   \dot{\rho}_m^\alpha = \dfrac{|\dot{\gamma}^\alpha|}{b_s \Lambda_s^\alpha} - \dfrac{2 \hat{d}^\alpha}{b_s} \rho_m^\alpha |\dot{\gamma^\alpha}|
   - \dfrac{2 \check{d}^\alpha}{b_s} \rho_m^\alpha |\dot{\gamma^\alpha}|,
   \label{mobile_edge_dislo_density}
  \end{align}
  and,
  \begin{align}
    \dot{\rho}_d^\alpha = \dfrac{2 \hat{d}^\alpha}{b_s} \rho_m^\alpha |\dot{\gamma^\alpha}|
   - \dfrac{2 \check{d}^\alpha}{b_s} \rho_d^\alpha |\dot{\gamma^\alpha}| - \rho_d^\alpha \dfrac{4 v_{climb}}{\hat{d}^\alpha - \check{d}^\alpha},
    \label{immobile_dipole_dislo_density}
   \end{align}
 respectively. $\Lambda_s^\alpha$ is the mean free path of dislocations.
 The maximum separation of the glide planes that allow dislocations gliding on them to form dipoles is $\hat{d^\alpha}$ 
 while two edge dislocations would get annihilated whenever they are any closer to each other than $\check{d^\alpha}$.  
 These distances are calculated to be,
 \begin{align}
  \hat{d^\alpha} = \dfrac{3 G b_s}{16\pi|\tau^\alpha|},
  \label{dist_dipol_formation}
 \end{align}
 and,
 \begin{align}
  \check{d^\alpha} = C_{anni} b_s,
  \label{dist_annihilation}
 \end{align}
 where, $C_{anni}$ is a fitting parameter.
 The dislocation climb velocity $v_{climb}$ is given by,
 \begin{align}
  v_{climb} = \dfrac{3 G D_0 \Omega}{2 \pi k_\text{B} T} \dfrac{1}{\left( \hat{d^\alpha} + \check{d^\alpha}\right)} \exp \left(- \dfrac{Q_c}{k_\text{B} T} \right),
  \label{velocity_dislo}
 \end{align}
 where $D_0$ is the self diffusion coefficient of the material in question, $\Omega$ and $Q_c$ are the activation volume and activation energy for climb, respectively.
 The mean free path $\Lambda_s^\alpha$ is defined as,
 \begin{align}
  \dfrac{1}{\Lambda_s^\alpha} = \dfrac{1}{d} + \dfrac{1}{\lambda_{slip}^\alpha},
  \label{def_mfp}
 \end{align}
 where $d$ is the grain size and $\lambda_{slip}^\alpha$ is the average distance traveled by a dislocation before it is stopped by forest dislocations, written as,
 \begin{align}
  \dfrac{1}{\lambda_{slip}^\alpha} = \dfrac{1}{i_{slip}}  \left[ \sum_{{\alpha'}=1}^{N_s} \xi_{\alpha \alpha'} \left( \rho_m^{\alpha'} + \rho_d^{\alpha'} \right)  \right]^{1/2},
  \label{lambda_slip}
 \end{align}
 where $i_{slip}$ is a fitting parameter.

 \section{Athermal mechanisms of EP }\label{athermal_mech}
 In this section we discuss the major theories explaining EP which do not rely on thermal softening 
 of the material due to Joule heating. We begin with a discussion of the theories by Molotskii and co-workers which 
 involve de-pinning of dislocations from obstacles during electropulsing.
 We follow that up with a review of the theory of electron-wind force assisted dislocation glide as put forth by Conrad and co-workers.  
 While considering each of these mechanisms we will also lay out the extensions 
 to the crystal plasticity model required to implement them and attempt a theoretical analysis of the reductions in flow stress caused by each of them 
 wherever possible. 
 
 \subsection{Paramagnetic de-pinning of dislocations}\label{dep_dislo}
Molotskii and Fleurov~\cite{Molotskii1995} in their seminal paper suggested that de-pinning of dislocations from paramagnetic 
obstacles is the dominant softening mechanism during electrical pulsing. They demonstrate that the induced magnetic field due to electrical 
pulsing alters the electronic states in the obstacles and the dislocation cores. 
These modified electronic states result in a much lower probability of the dislocations 
being pinned by such obstacles. Forest dislocations qualify as paramagnetic obstacles \cite{Molotskii1995_2} 
and they constitute the largest fraction of short range obstacles encountered by dislocations in FCC materials.
Thus, any gain in plasticity due to a change in the pinning behaviour of such obstacles is likely to be very significant.

It must be noted that the mechanism of de-pinning of dislocations is not dependent on the sense
of the current density vector $\boldsymbol{j}$, but only on its magnitude $j=|\boldsymbol{j}|$.
So, in this section, whenever we use ``current density'' we mean its magnitude represented by $j$. 

In view of the crystal plasticity model we have described in Section~\ref{cpmodel}, reductions in flow stress due to de-pinning of dislocations 
can manifest through several terms in Eq.~\ref{Orowan_eq}. We will refer to them as ``sources'' of softening 
within the purview of the primary mechanism of de-pinning of dislocations from paramagnetic obstacles.
We investigate each of the sources in the following sub-sections.

\subsubsection{Effect of a change in $\tau_{sol}$}
As proposed by Molotskii and co-workers~\cite{Molotskii1995, Molotskii2000}, during the application of an electrical pulse,
the pinning of dislocations by short range obstacles are weakened.
This implies that the 
inter-obstacle spacing $l_c(0)$ while no current is passed
increases to a new value $l_{c}(j)$ under pulsing. 
The inter-obstacle distance ($l_c$) changes as a function of the 
imposed current density ($j$) as~\cite{Molotskii1995},
\begin{align}
 l_c(j) = l_c(0) \left( 1 + \dfrac{j^2}{j_0^2} \right),
 \label{mol_pin_j}
\end{align}
where, $j_0$ is a characteristic current density magnitude which corresponds 
to the magnitude of current density $j$ at which the 
EP is typically observed for a particular material. 

The change in $l_c$ will affect the parameter $\tau_{sol}$ in Eq.~\ref{Orowan_eq}.
$\tau_{sol}$ represents the short range resistance experienced by an average dislocation segment.
Following the analysis in \cite{Kocks1975},
$\tau_{sol}$ scales inversely as the separation between the short range obstacles.
In other words, 
\begin{align}
 \tau_{sol} \propto \dfrac{1}{l_c}.
 \label{tsol_with_lc}
\end{align}
Combining Eqs.~\ref{mol_pin_j} and~\ref{tsol_with_lc} we can write $\tau_{sol}$
as a function of $j$ as,
\begin{align}
 \tau_{sol}(j) = \dfrac{\tau_{sol}(0)}{\left( 1 + \dfrac{j^2}{j_0^2} \right)}.
 \label{tau_sol_j}
\end{align} 
Under electropulsing the shear rate on any particular slip system $\alpha$ can be written as,
\begin{align}
 \dot{\gamma^\alpha}(j) = \rho_m^\alpha b_s v_0 \exp\left[ - \dfrac{Q_s}{k_\text{B} T} \left\{ 1 - \left( \dfrac{\tau_{eff}^\alpha(j)}{\tau_{sol}(j)}\right)^p \right\}^q \right] \textrm{sign}(\tau^\alpha).
 \label{Orowan_mol_tau_sol}
\end{align}
The lowering of $\tau_{sol}$ represents lowering of the short range obstacle strength. 
Thus, under pulsing, similar shear rates ($\dot{\gamma^\alpha}$) 
can be maintained by smaller effective resolved shear stresses 
$\tau_{eff}^\alpha$ and consequently  by $\tau^{\alpha}$ which explain
the reductions in flow stress on the level of individual slip systems. 
Thus, the dependence of $\tau_{sol}$ on $j$ (Eq.~\ref{tau_sol_j}), 
leads to $\tau_{eff}^\alpha$ being a function of $j$ 
in Eq.~\ref{Orowan_mol_tau_sol}.

Molotskii and Fleurov\cite{Molotskii1995} also perform an analysis of the estimated stress drop due to such a softening mechanism.
While our formulation of the electroplastic phenomena is on the slip system level, the analysis performed in~\cite{Molotskii1995}
is for the bulk polycrystalline sample.
In order to replicate a similar analysis employing our formulation,
we relate the plastic behaviour of the bulk sample
to that of a single grain employing the concept of Taylor factor ($M$),
\begin{align}
 &\dot{\epsilon}(0) = \dfrac{1}{M}\sum_{\alpha=1}^{N_s} \dot{\gamma^\alpha}(0) \nonumber \\
                &= \dfrac{1}{M}\sum_{\alpha=1}^{N_s} \Bigg( \rho_m^\alpha b_s v_0 \cdot \nonumber \\ 
                &\exp\left[ - \dfrac{Q_s}{k_\text{B} T} \left\{ 1 - {\left( \dfrac{(\sigma_{eff}(0)/M)}{\tau_{sol}(0)} \right) }^p 
   \right\}^q \right] \Bigg) \textrm{sign}(\tau^\alpha),
   \label{Orowan_no_pulsing_bulk}
\end{align}
where, $\alpha$ runs over all the slip systems in single grain. In Eq.~\ref{Orowan_no_pulsing_bulk} $\dot{\epsilon}$ is the imposed bulk strain rate and remains constant
regardless of whether the system is pulsed or not and $\sigma_{eff}$ is the effective normal stress along the loading direction and can be thought of as
the difference between the applied stress ($\sigma_{appl}$) and the long-range resistance ($\sigma_{pass}$) which are related to the corresponding quantities
$\tau^\alpha$ and $\tau_{pass}^\alpha$, respectively, for a single slip system.
Eq.~\ref{Orowan_no_pulsing_bulk}, can be modified for an electropulsed sample in the following manner as,
\begin{align}
 &\dot{\epsilon}(j) = \dfrac{1}{M}\sum_\alpha \dot{\gamma^\alpha}(j) \nonumber \\
                &= \dfrac{1}{M}\sum_\alpha \Bigg( \rho_m^\alpha b_s v_0 \cdot  \nonumber \\
                &\exp\left[ - \dfrac{Q_s}{k_\text{B} T} \left\{ 1 - {\left( \dfrac{(\sigma_{eff}(j)/M)}{\tau_{sol}(j)} \right) }^p 
   \right\}^q \right] \Bigg) \textrm{sign}(\tau^\alpha).
   \label{Orowan_pulsing_bulk}
\end{align}
We divide Eq.~\ref{Orowan_pulsing_bulk} by Eq.~\ref{Orowan_no_pulsing_bulk} 
and impose $\dot{\epsilon}(j) = \dot{\epsilon}(0)$, to write,
\begin{align}
   &\sum_\alpha \rho_m^\alpha  \exp\left[ - \dfrac{Q_s}{k_\text{B} T} \left\{ 1 - {\left( \dfrac{(\sigma_{eff}(0)/M)}{\tau_{sol}(0)} \right) }^p 
   \right\}^q \right] \textrm{sign}(\tau^\alpha) \nonumber \\  &=  \sum_\alpha \rho_m^\alpha \exp\left[ - \dfrac{Q_s}{k_\text{B} T} \left\{ 1 - {\left( \dfrac{(\sigma_{eff}(j)/M)}{\tau_{sol}(j)} \right) }^p 
   \right\}^q \right] \textrm{sign}(\tau^\alpha).
   \label{eq_Orowan_mol}
\end{align}
Noting that the arguments to the exponentials are not functions of $\alpha$, we can simplify Eq.~\ref{eq_Orowan_mol} to get,
\begin{align}
\dfrac{\sigma_{eff}(0)}{\tau_{sol}(0)} = \dfrac{\sigma_{eff}(j)}{\tau_{sol}(j)},
\label{str_drop_mol_ratio}
\end{align}
and using Eq.~\ref{tau_sol_j} this can be written as,
\begin{align}
 \Delta \sigma_{eff} = \sigma_{eff}(0) - \sigma_{eff}(j) = \sigma_{eff}(0) \dfrac{j^2}{j_0^2 + j^2}.
 \label{str_drop_mol_fin_eff}
\end{align}
$\Delta \sigma_{eff}$ is the difference in effective stresses recorded before and after electropulsing 
and we can relate it to the 
applied stress ($\sigma_{appl}$) by assuming the averaged long range elastic stress fields 
to remain constant ($\sigma_{pass}$) when $\tau_{sol}$ is modified due to electropulsing. 
So, in terms of the applied stresses,
\begin{align}
  \Delta \sigma_{appl} = \Delta \sigma_{eff} = \sigma_{eff}(0) \dfrac{j^2}{j_0^2 + j^2}.
 \label{str_drop_mol_fin}
\end{align}
The above equation reveals interesting trends for very low and high values of $j$. For $j<<j_0$,
\begin{align}
 \Delta \sigma_{appl} \approx  \sigma_{eff}(0) \dfrac{j^2}{j_0^2},  
 \label{str_drop_mol_in_para}
\end{align}
which reveals a parabolic dependence on $j$. When $j>>j_0$, we get,
\begin{align}
  \Delta \sigma_{appl} \approx  \sigma_{eff}(0),  
  \label{str_drop_late_const}
\end{align}
and so the stress drop saturates. The variation of $\Delta \sigma_{appl}$ as a function of $(j/j_0)$ 
as predicted by Eq.~\ref{str_drop_mol_fin} has a point of inflection where the curvature $(\partial^2 \Delta \sigma_{appl}/\partial j^2)$ changes 
from positive to negative and it happens at $(j/j_0)=1/\sqrt{3} = 0.577$, which marks a transition between the regimes denoted by Eqs.~\ref{str_drop_mol_in_para}
and~\ref{str_drop_late_const}.

It is also important to take note of the assumptions made to arrive at Eq.~\ref{str_drop_mol_fin}.
The one central assumption of the analysis is that the effective resolved shear stresses $\tau_{eff}^\alpha$
are the same for all the slip systems ($\alpha$). This assumption allows us to introduce the Taylor factor($M$)
in Eq.~\ref{Orowan_no_pulsing_bulk}. For crystal plasticity simulations of bulk polycrystalline materials an
equality of $\tau_{eff}^\alpha$ for all the slip systems for any particular grain is uncommon and there are considerable 
differences across the slip systems over all the grains. 
Also, it has even been observed that the number 
of active slip systems ($|\tau^\alpha| > \tau_{pass}^\alpha$) may vary from $3$ to $9$ 
per grain in a polycrystalline sample and all the grains 
do not deform similarly as conjectured by Taylor's theory~\cite{Taylor1938}.
So, in view of these differences with crystal plasticity implementations, 
it is reasonable to expect differences between the predictions from simulations and the simplified analysis.

While considering a lowered short-range obstacle strength, 
Molotskii and co-workers ignored other potential ramifications of the de-pinning of 
dislocations from paramagnetic obstacles. In the next few subsections, we will explore other possible consequences
of dislocation de-pinning within the crystal plasticity framework we have introduced.

\subsubsection{Effect of a change in $\lambda_{slip}^\alpha$}\label{sec_lambda_slip}
Another possible effect of the de-pinning of dislocations is an increase of the 
distance traveled by the dislocation before being trapped by the forest dislocations ($\lambda_{slip}^\alpha$).
$\lambda_{slip}^\alpha$ should be scaled by the same factor $(1 + j^2/j_0^2 )$ which depicts the scaling of the 
distance between the trapping points due to the forest dislocations. Hence, we can write from Eq.~\ref{lambda_slip},  
\begin{align}
  \lambda_{slip}^\alpha(j) =  \lambda_{slip}^\alpha(0) \left( 1 + \dfrac{j^2}{j_0^2} \right) 
  \label{lambda_slip_pulse}
 \end{align}
The effect of change in this length scale on the applied stress is complicated. 
It alters the evolution of the dislocation densities through Eq.~\ref{mobile_edge_dislo_density}
which in turn can affect the passing stress $\tau_{pass}^\alpha$ as given in Eq.~\ref{tau_pass_def}. 
As $\lambda_{slip}^\alpha$ does not explicitly appear in the Orowan equation an 
analytical expression relating it to the reductions in flow stress is intractable.
However, it is worthwhile 
to develop an intuitive understanding of the kind of changes prompted in $\rho_m^\alpha$ and $\tau_{pass}^\alpha$
due to a change in $\lambda_{slip}^\alpha$. 
From Eqs.~\ref{mobile_edge_dislo_density} and~\ref{def_mfp} it is clear that
$\rho_m^\alpha$ would increase at a slower rate when $\lambda_{slip}^\alpha$ is larger.
Under the assumption that $\dot{\epsilon(0)} = \dot{\epsilon(j)}$, implies 
$\dot{\gamma^\alpha}(0) = \dot{\gamma^\alpha}(j)$,  a smaller $\rho_m^\alpha$ 
necessitates a larger $\tau_{eff}^\alpha$ to maintain a constant $\dot{\gamma^\alpha}$.  
A higher $\tau_{eff}^\alpha$ can be achieved either by lowering of $\tau_{pass}^\alpha$
or by a higher resolved shear stress ($\tau^\alpha$). The former can lead to flow softening
while the latter can lead to a stress rise during electropulsing and the net change is 
a combined effect of the two. Thus, there is a possibility of a rise in flow stress 
instead of a drop due to an increase of $\lambda_{slip}^\alpha$.   

\subsubsection{Effect of a change in $v_0$}
From the expression of the Orowan equation presented in Eq.~\ref{Orowan_eq},
the velocity of dislocations can be expressed as a product of the 
velocity prefactor $v_0$ and an Arrhenius term,
given as,
\begin{align}
v = v_0 \exp\left[ - \dfrac{Q_s}{k_\text{B} T} \left\{ 1 - \left( \dfrac{\tau_{eff}^\alpha}{\tau_{sol}}\right)^p \right\}^q \right],
\label{dislo_vel}
\end{align}
where, 
\begin{align}
 v_0 = \nu_G \,  d_s,
 \label{velocity_def}
\end{align}
where $\nu_G$ is the jump frequency of the dislocations and $d_s$ is the distance moved forward per successful thermal activation event.
Granato et al.,~\cite{Granato1964} suggest that $\nu_G$ is independent of the free length of the dislocations between two obstacles
and so should remain constant
as the odds of pinning by obstacles lower while electropulsing. As an obstacle is overcome by 
thermal activation, the freed dislocation segment glides until it encounters another obstacle, after which the entire process 
of thermal activation is repeated. From that argument, we have  $d_s = l_c$. 
Thus, under electropulsing, it is reasonable to expect that dislocations would glide 
larger distances as pinning is less probable. 
Using the variation of $l_c(j)$ as given by Eq.~\ref{mol_pin_j} 
we can write the corresponding scaling relationship for $d_s$ as,
\begin{align}
 d_s(j) = d_s(0) \left( 1 + \dfrac{j^2}{j_0^2} \right).
 \label{scaling_for_d}
\end{align}
Combining Eqs.~\ref{velocity_def} and~\ref{scaling_for_d} we get a 
scaling relationship for $v_0$ as given by,
\begin{align}
 v_0 (j) = v_0(0) \left( 1 + \dfrac{j^2}{j_0^2} \right).
\end{align}

An analysis similar to that done for the reductions in flow stress due to $\tau_{sol}$ when carried out for this case
results in an expression which reads,
\begin{align}
 \Delta \sigma_{appl} = \Delta \sigma_{eff} = M \tau_{sol} \left(\dfrac{k_\text{B} T}{Q_s} \right) \log \left( 1 + \dfrac{j^2}{j_0^2} \right),
 \label{str_drop_v0}
\end{align}
where, we have assumed $\sigma_{pass}$ to be unaffected by a change in $v_0$ during pulsing.
For situations where $j/j_0 << 1$, Eq.~\ref{str_drop_v0} can be approximated as,
\begin{align}
 \Delta \sigma_{appl} = \Delta \sigma_{eff} = M \tau_{sol} \left(\dfrac{k_\text{B} T}{Q_s} \right) \dfrac{j^2}{j_0^2}.
 \label{str_drop_v0_shortened}
\end{align}

\subsubsection{Effect of a change in passing stress $\tau_{pass}^\alpha$}
Molotskii and Fleurov~\cite{Molotskii1996} argue that as the dislocations are de-pinned from short range obstacles,
they have a larger free length and hence have a larger geometrical freedom to rotate and re-orient 
to the long range elastic stress fields of other dislocations. Thus, under electropulsing, dislocations can achieve 
configurations which minimize the strain energy more effectively compared to the situation where no 
electrical pulses are applied. This implies that larger stresses are needed 
to force dislocation motion while being pulsed and is an anomaly considering that
all other mechanisms induce softening of the material. 
In the context of the crystal plasticity model 
described in the paper, $\tau_{pass}^\alpha$ represents the resistance to dislocation motion 
from the long range strain fields and it should be modified to describe this particular 
phenomenon. Following the analysis of Molotskii and Fleurov~\cite{Molotskii1996} we can modify Eq.~\ref{tau_pass_def} as,
\begin{align}
 \tau_{pass}^\alpha = \delta G b_s \left[ \sum_{{\alpha'}=1}^{N_s} \xi_{\alpha \alpha'} \left( \rho_m^{\alpha'} + \rho_d^{\alpha'} \right)  \right]^{1/2},
 \label{mod_passing_stress}
\end{align}
where, a factor $\delta$ is introduced to account for the enhanced 
elastic interaction. $\delta$ is conjectured to have a form given by
the following expansion~\cite{Molotskii1996},
\begin{align}
 \delta &=  1 + \beta \left(\dfrac{l_c(j) - l_c(0)}{l_c(0)}\right) + \cdots , \nonumber \\
        &= 1 + \beta \dfrac{j^2}{j_0^2},
 \label{delta_def}
\end{align}
restricting consideration to the first order terms only. In Eq.~\ref{delta_def}
$\beta$ is a constant determined by fitting and we have made use of Eq.~\ref{mol_pin_j}
to arrive to the final form.

We can derive an expression for reductions in flow stress (or increases in flow stress in this case) $\Delta \sigma_{appl}$ corresponding to the 
enhanced work hardening due to electropulsing. Following the approach undertaken to derive Eq.~\ref{str_drop_mol_ratio},
a corresponding relationship for this case is written as,
\begin{align}
  \sigma_{eff}(j) = \sigma_{eff}(0),
  \label{str_drop_tau_pass_change}
\end{align}
which leads to,
\begin{align}
 \Delta \sigma_{appl} = \Delta \sigma_{pass} = - \sigma_{pass}(0) \beta \dfrac{j^2}{j_0^2},
 \label{str_drop_tau_pass}
\end{align}
where we have assumed $M \tau_{pass}^\alpha = \sigma_{pass}$, $M$ being the Taylor factor. 
The negative sign in the RHS of Eq.~\ref{str_drop_tau_pass} 
confirms that the applied stresses would need to be increased when such an effect
is operative. 

It is important that to note $\beta$ can be determined by fitting Eq.~\ref{mod_passing_stress} to the 
variation of yield stresses in the material as a function of the imposed current density $j$~\cite{Molotskii1996}.
The yield stress of the material corresponds to the value of $\sigma_{pass}$ when $\sigma_{applied}$ 
overcomes $\sigma_{pass}$ for the first time. Such a quantity is not modified by any other softening 
mechanism operative due to de-pinning of dislocations and hence is suitable for determining $\beta$.

We have considered all possible manifestations of the phenomenon of de-pinning of dislocations during electropulsing.
Also, we have presented extensions to the crystal plasticity model which will allow us to simulate the effects of these mechanisms
in the later sections. We will discuss the theory of electron-wind force assisted dislocation motion next.

\subsection{Electron-wind force: Conrad and co-workers}
Conrad et al.,~\cite{Conrad1990} proposed a theory where the electrons drifting under the application of an electric field exert a force
on the dislocations. This is known as ``electron-wind'' force and its expression derived
by considering the scatter of electrons by dislocations can be stated as,
\begin{align}
  \boldsymbol{F_{ew}} = \dfrac{\rho_D}{N_D} e n_e  \boldsymbol{j} ,
 \label{conrad_few_scatter}
\end{align}
where, $e$ is the electronic charge, $n_e$ is the density of free electrons, $\boldsymbol{j}$ is the current density and $\boldsymbol{F_{ew}}$ is the 
force per unit length of dislocations. In the crystal plasticity model described, plasticity is governed 
by the motion of the pure edge dislocations which move parallel to themselves along the slip direction $\boldsymbol{m^\alpha}$
in the particular slip systems ($\alpha$). In that case the component of the electron-wind force   
along the direction of the dislocation motion is given by $\boldsymbol{F_{ew}} \cdot \boldsymbol{m^\alpha}$.
Assuming an equal fraction of positive and negatively signed dislocations in the material, 
it is apparent that if the electron-wind force aids the gliding dislocations of one particular sign then 
it should impede the motion of those belonging to the other sign. This is also pointed out by Molotskii and Fleurov\cite{Molotskii1995} 
and it implies that the electron-wind force on dislocations is at best a second order effect.   

Modeling the effects of the electron-wind force requires 
an alternate form of the Orowan equation (Eq.~\ref{Orowan_eq}) written as,
\begin{align}
  \dot{\gamma^\alpha} = \rho_m^\alpha b_s v_0 \exp\left[ - \left(\dfrac{Q_s - \tau_{eff}^\alpha V}{k_\text{B} T} \right) \right] \textrm{sign}(\tau^\alpha),
  \label{Orowan_eq_mod}
\end{align}
where, the argument of the exponential has been written in an equivalent form assuming the following relationship, 
\begin{align}
\Delta G = Q_s -  \tau_{eff}^\alpha V^\alpha = Q_s \left\{ 1 - \left( \dfrac{\tau_{eff}^\alpha}{\tau_{sol}}\right)^p \right\}^q, 
\label{exponential_argument}
\end{align}
with the Gibbs free-energy of activation denoted by $\Delta G$.
The new parameter which appears in the modified Orowan equation (Eq.~\ref{Orowan_eq_mod}) is the activation volume 
of slip denoted by $V^\alpha$. The new form of the Orowan equation is necessitated to simplify the introduction of the 
electron-wind force. Following Conrad\cite{Conrad1990}, the effect of the electron-wind force can be modeled as,
\begin{align}
  \dot{\gamma^\alpha} &= \dfrac{1}{2}\rho_m^\alpha b_s v_0 \Bigg\{ \exp\left[ - \left(\dfrac{Q_s - \tau_{eff}^\alpha V^\alpha - (\boldsymbol{F_{ew}} \cdot \boldsymbol{m^\alpha}) A^\alpha}{k_\text{B} T} \right) \right] + \nonumber \\
                       & \exp\left[ - \left(\dfrac{Q_s - \tau_{eff}^\alpha V^\alpha + (\boldsymbol{F_{ew}} \cdot \boldsymbol{m^\alpha}) A^\alpha}{k_\text{B} T} \right) \right] \Bigg\}\textrm{sign}(\tau^\alpha),
  \label{Orowan_eq_elec_wind}
\end{align}
where $A^\alpha$ denotes the activation area which is related to the activation volume ($V$) as, $V^\alpha = A^\alpha b_s$.
Eq.~\ref{Orowan_eq_elec_wind} treats the electron-wind as an additional force 
on the dislocations similar to that exerted by $\tau_{eff}^\alpha$. 
The differential effect 
of the electron-wind force on dislocations
of either sign manifests in Eq.~\ref{Orowan_eq_elec_wind} 
through opposite signs of the electron-wind force
$(\boldsymbol{F_{ew}} \cdot \boldsymbol{m^\alpha})$
for the dislocation densities of either signs $(\rho_m^\alpha/2)$. 
The usefulness of the form of 
Eq.~\ref{Orowan_eq_mod} is evident here as it 
allows the electron-wind force to be treated in a manner 
akin to the effective stress $\tau_{eff}^\alpha$. 
Eq.~\ref{Orowan_eq_elec_wind} can be simplified to obtain,
\begin{align}
   &\dot{\gamma^\alpha} = \rho_m^\alpha b_s v_0 \exp\left[ - \left(\dfrac{Q_s - \tau_{eff}^\alpha(j) V^\alpha}{k_\text{B} T} \right) \right] \nonumber \\ 
   &\cosh \left(\dfrac{(\boldsymbol{F_{ew}} \cdot \boldsymbol{m^\alpha}) A}{k_\text{B} T} \right)\textrm{sign}(\tau^\alpha).
\end{align}
We are now in a position to revert back to the form of the Orowan equation as given by Eq.~\ref{Orowan_eq}
to finally write,
\begin{align}
 &\dot{\gamma^\alpha} = \nonumber \\
                     &\rho_m^\alpha b_s v_0 \exp\left[ - \dfrac{Q_s}{k_\text{B} T} \left\{ 1 - \left( \dfrac{\tau_{eff}^\alpha(j)}{\tau_{sol}}\right)^p \right\}^q \right]
                     \nonumber \\
 &\cosh \left(\dfrac{(\boldsymbol{F_{ew}} \cdot \boldsymbol{m^\alpha})A}{k_\text{B} T} \right) \textrm{sign}(\tau^\alpha),
 \label{Orowan_eq_elec_wind_final}
\end{align}
which indicates that the electron-wind force introduces the factor, $\cosh \left((\boldsymbol{(F_{ew}} \cdot \boldsymbol{m^\alpha}) A)/(k_\text{B} T) \right) $,
to the Orowan equation. For a situation where, $(( \boldsymbol{F_{ew}} \cdot \boldsymbol{m^\alpha})A/(k_\text{B} T)) << 1$, a Taylor-Series expansion 
writes as,
\begin{align}
 \cosh \left(\dfrac{(\boldsymbol{F_{ew}} \cdot \boldsymbol{m^\alpha})A}{k_\text{B} T} \right)=  
 1 + \dfrac{1}{2!}\left( \dfrac{(\boldsymbol{F_{ew}} \cdot \boldsymbol{m^\alpha})A}{k_\text{B} T} \right)^2 + \cdots.
 \label{cosh_expansion}
\end{align}
So, this indicates that in the event the electron-wind force is small, the factor introduced by it in the Orowan equation 
is of the second order.

An analytical approach to determine the dependence of reductions in flow stress on the current density ($j$) is not feasible
due to the dependence of the electron-wind force on the direction of dislocation motion on individual slip planes. 
Such a dependence complicates deriving an analytical expression of the average electron-wind force over the entire bulk sample.

Conrad and co-workers~\cite{Conrad1990} also claimed that the activation area $A^\alpha$ changes due to electropulsing.
The activation area ($A^\alpha$) can be defined as~\cite{Molotskii1995},
\begin{align}
A^\alpha = -\dfrac{1}{b_s}\dfrac{\partial \Delta G}{\partial \tau_{eff}^\alpha},
\label{activ_energy}
\end{align}
and using the last equality of Eq.~\ref{exponential_argument} in Eq.~\ref{activ_energy}, we get,
\begin{align}
 A^\alpha = \dfrac{Q_s}{\tau_{sol} b_s} pq\left\{ 1 - \left(\dfrac{\tau_{eff}^\alpha}{\tau_{sol}}\right)^p  \right\}^{(q-1)} \left(\dfrac{\tau_{eff}^\alpha}{\tau_{sol}}\right)^{(p-1)}.
 \label{activation_area}
\end{align}
From Eq.~\ref{activation_area}, it is clear that $A^\alpha$ increases under electropulsing as $\tau_{sol}$ is scaled down by a factor of $\left(1 + \dfrac{j^2}{j_0^2}\right)$
following our implementation of Molotskii's theory~\cite{Molotskii1995}. So, henceforth in our discussion we do not consider a change in $A^\alpha$
explicitly as such an effect is already included in the mechanism of a change in $\tau_{sol}$ causing softening. 

\section{Thermal softening due to Joule-heating}
Another source of the reductions in flow stress during electropulsing is conjectured to be the thermal softening of the material due to 
Joule heating. In order to examine this particular source, we first solve the thermal conduction equation to determine the temperature ($T$), 
\begin{align}
 \rho C_p \dot{T} = \nabla \cdot \left( \boldsymbol{K} \nabla T \right) + \dot{Q}, 
 \label{therm_cond}
\end{align}
where, $\boldsymbol{K}$ is the thermal conductivity, $\rho$ is the density of the material, $C_p$ is the specific heat capacity, and 
$\dot{Q}$ is the source term per unit volume. The heat source term is computed as,
\begin{align}
 \dot{Q} &= \boldsymbol{j} \cdot \boldsymbol{E}, \nonumber \\
         &= \boldsymbol{j} \cdot \left[{\boldsymbol{\sigma_{el}}}^{-1}\boldsymbol{j}\right],
  \label{th_source_def}       
\end{align}
 where, we have used $\boldsymbol{j}=\boldsymbol{\sigma_{el}}\boldsymbol{E}$; $\boldsymbol{E}$ is the electric field vector and $\boldsymbol{\sigma_{el}}$ 
 is the electrical conductivity tensor. 
 The evolution of $T$ impacts the shear rates ($\dot{\gamma^\alpha}$) as given by,
 \begin{align}
   \dot{\gamma^\alpha} = \rho_m^\alpha b_s v_0 \exp\left[ - \dfrac{Q_s}{k_\text{B} T(j)} \left\{ 1 - \left( \dfrac{\tau_{eff}^\alpha(j)}{\tau_{sol}}\right)^p \right\}^q \right] \textrm{sign}(\tau^\alpha),
 \end{align}
 where $T$ is now a function of the imposed current density $j$. An analysis to determine the reductions in flow stress from thermal softening can be carried out 
 in a manner similar to that done for the situation where $\tau_{sol}$ changes. In order to do that, we first relate the change in temperature 
 to the imposed current density ($j$). For a single phase material which is isotropic and homogeneous in all properties, tensors $\boldsymbol{K}$, 
 $\boldsymbol{\sigma_{el}}$
 can be reduced to scalars $K$ and $\sigma_{el}$. The homogeneity of $\sigma_{el}$ implies that the 
 source term $\dot{Q}$ is also homogeneous, i.e., every point in the domain experience the same heat source.  
 Under such approximations $\nabla T = 0 $ as there is no reason for the $T$ field to be inhomogeneous. This 
 simplifies Eq.~\ref{therm_cond} into,
 \begin{align}
  \rho C_p \dot{T} =  \dot{Q}. 
 \label{therm_cond_simp}
 \end{align}
 Solving the above equation yields,
 \begin{align}
  T(j) = T(0) + P j^2,
  \label{temp_sim}
 \end{align}
 where, $P=\Delta t/\sigma_{el}\rho C_p$ with $\Delta t$ being the pulse duration. 
 Temperatures at the beginning and at the end of the pulse are denoted by $T(0)$ and $T(j)$.
 We have ignored cooling of the sample to maintain simplicity of the formulation.
 
 We resume our analysis to determine the dependence of reductions in flow stress on the current density $j$, 
 and re-tracing the steps employed to derive Eq.~\ref{str_drop_mol_ratio} we obtain, 
 \begin{align}
  \left[\dfrac{1}{T(0)} - \dfrac{1}{T(j)} \right] = \dfrac{1}{M \tau_{sol}}\left(\dfrac{\sigma_{eff}(0)}{T(0)} - \dfrac{\sigma_{eff}(j)}{T(j)} \right).
  \label{str_drp_therm_sof_ratio}
 \end{align}
 Now, employing Eq.~\ref{temp_sim} in the above equation we get, 
 \begin{align}
  \Delta \sigma_{appl} = \Delta \sigma_{eff} = \dfrac{P}{T(0)}j^2 \left[M \tau_{sol} - \sigma_{eff}(0)\right],
  \label{str_drp_therm_sof}
 \end{align}
  which again assumes constancy of $\sigma_{pass}$ before and after pulsing. 
  Thus, the reductions in flow stress have a quadratic dependence on the current density $j$ 
  when thermal softening is the operative mechanism. 
  It must be noted that in the presented analysis we have assumed that dislocation climb has not played a role.
  We will comment on the validity of this assumption based on the temperature rises seen during our simulations of electropulsing.
   
  In the previous sections, we have discussed in detail the crystal plasticity model for simulating EP.
  We have also provided some analytical expressions relating the stress drop to the current density. 
  At this point we can present the form of the Orowan equation at the slip system level which displays 
  contributions from all the mechanisms of EP,
  \begin{align}
   \dot{\gamma^\alpha}(j) = \rho_m^\alpha b_s v_0(j) 
   \exp\left[ - \dfrac{Q_s}{k_\text{B} T(j)} \left\{ 1 - \left( \dfrac{\tau_{eff}^\alpha(j)}{\tau_{sol}(j)}\right)^p \right\}^q \right] \nonumber \\
   \cosh \left(\dfrac{(\boldsymbol{F_{ew}} \cdot \boldsymbol{m^\alpha}) A}{k_\text{B} T} \right)  \textrm{sign}(\tau^\alpha),
   \label{Orowan_ep_complete}
  \end{align}
  where, $\tau_{eff}^\alpha(j) = |\tau^\alpha(j)| - \tau_{pass}^\alpha$, when $|\tau^\alpha(j)| > \tau_{pass}^\alpha$
  and $\tau_{eff}^\alpha(j)=0$, otherwise.
  
  It should be mentioned at this point that we have not considered the skin, pinch, and magnetostriction effects as 
  possible contributors to the phenomenon of EP as their contributions have been established to be small~\cite{Okazaki1980,Sprecher1986}
  compared to the mechanisms under discussion in this paper. However, thermal expansion due to Joule-heating is recognized 
  to have a bigger contribution compared to skin, pinch and magnetostriction effects~\cite{Sprecher1986}.
  But understandably such an effect is restricted only to tensile tests, and for the compression tests simulated in our paper,
  thermal expansion can lead to an increase of the flow stress. Thus, due to the lack of generality of the impact of thermal expansion
  on the flow stress, we have excluded it from our consideration.  
    
  In the following 
  section we report simulations of EP through which we attempt to understand the contribution 
  of individual mechanisms to the electroplastic effect.
  
  \section{Results}
  Before presenting the results of our simulations on the electroplastic effect, we would like to discuss a few details about 
  the numerical implementation and the solution technique. We implement the dislocation density based crystal plasticity model
  in the open-source crystal plasticity software DAMASK~\cite{Roters2019} and perform representative volume element 
  (RVE) simulations of uniaxial compression of polycrystalline samples using the spectral solver~\cite{Eisenlohr2013, Shanthraj2015}.
  
\begin{figure}[!htbp]
  \begin{center}
   \subfigure[]{\includegraphics[width=0.8\linewidth]{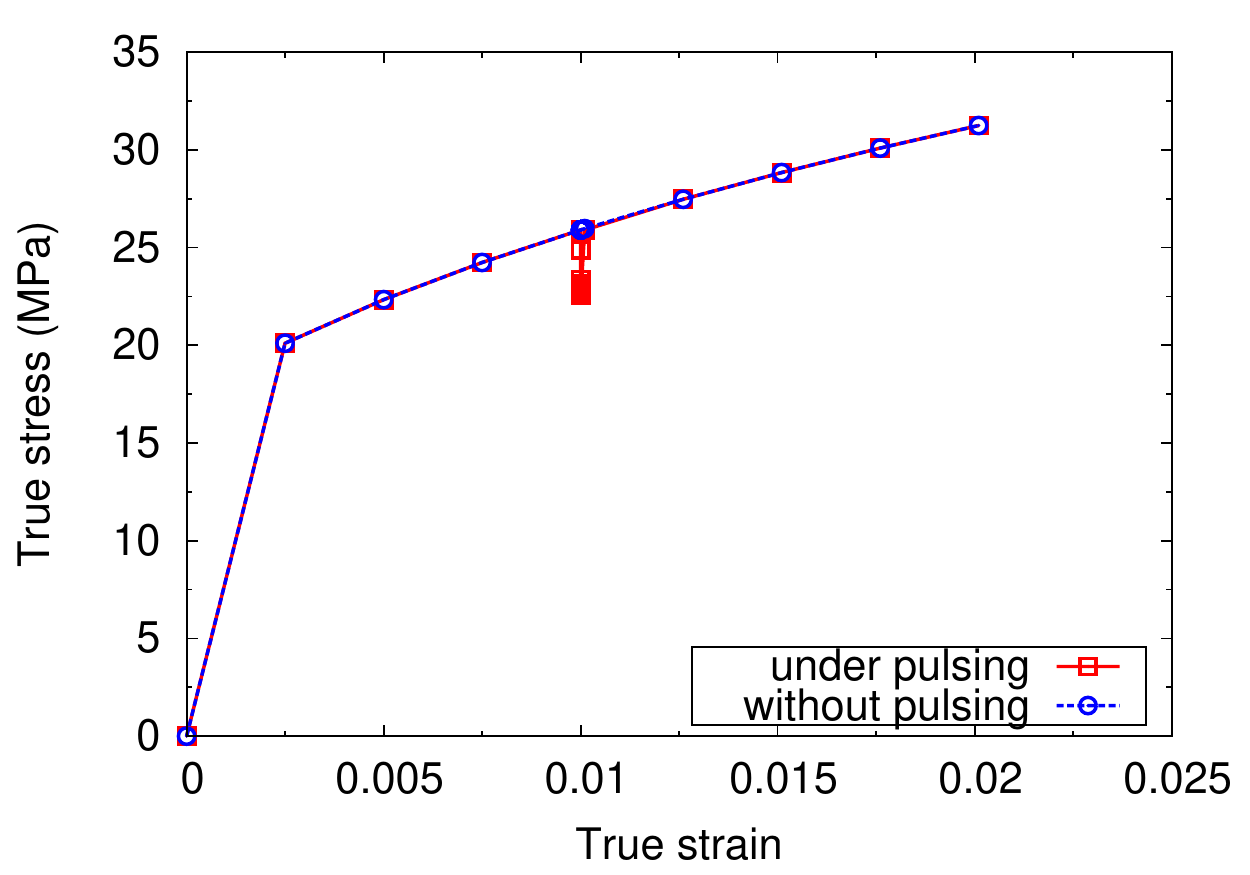}
    \label{flow_curve_tau_sol}}
   \subfigure[]{\includegraphics[width=0.8\linewidth]{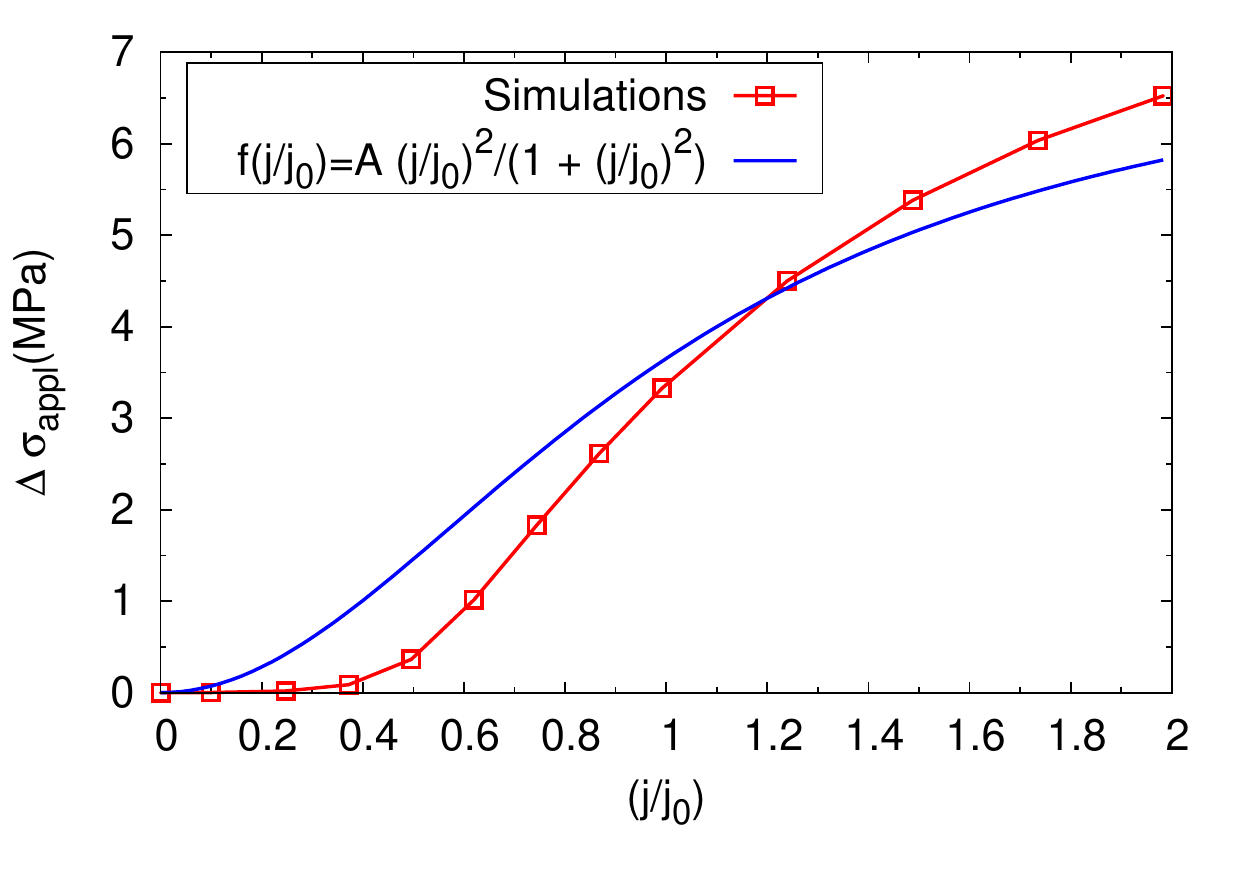}
   \label{str_drp_tau_sol_func_j}}
   \subfigure[]{\includegraphics[width=0.8\linewidth]{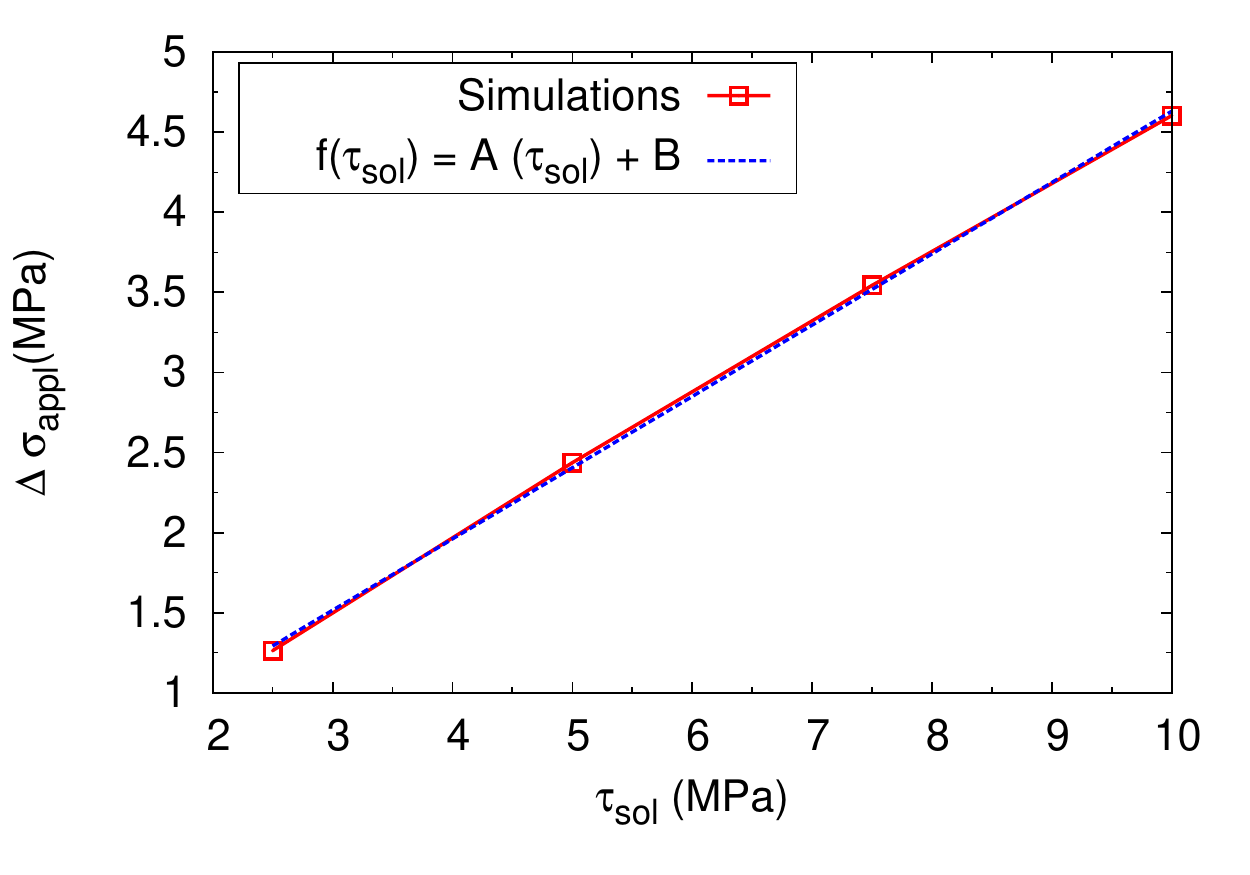}
   \label{str_drp_tau_sol_func_tau_sol}}
  \end{center}
  \caption{Figures demonstrating different aspects of the reductions in flow stress due to a change in $\tau_{sol}$. 
  (a) Flow curves with and without electropulsing during a compression test.
  The imposed strain rate is $\dot{\epsilon} = 1e-03 \text{s}^{-1}$
  and the pulse is applied at a strain of $0.01$. 
  The imposed pulse corresponds to $j/j_0 \approx 1$ 
  and is applied for a total time of $60 \, \mu \text{s}$.
  (b) Reductions in flow stress as a function of $(j/j_0)$.
  (c) reductions in flow stress as a function of $\tau_{sol}$ at $j=j_0$.
      The constants A and B in the figure legends are determined by fitting
      to the simulation data. 
  For (b) and (c) the loading and pulsing details are the same as in (a).  }
  \label{fig_drop_tau_sol}
 \end{figure}

  Regarding the choice of the crystal plasticity parameters, we must reiterate that our objective is not
  to simulate the electroplastic effect observed for any particular 
  material, but to explore the characteristics of each of the softening mechanisms of EP.  
  In that regard, we work with typical values of different parameters which are representative of a generic FCC material.
  We present the values of the parameters related to dislocation glide and climb in Table~\ref{tab_param_dislo}, while parameters
  related to the various mechanisms of EP are mentioned in Table~\ref{tab_param_ep}. The electrical and thermal properties 
  of the material are mentioned in Table~\ref{tab_param_el_mech_prop}. 
  In experiments of EP, the pulses are usually applied for 
  $\approx 100 \, \mu \text{s}$ while their magnitudes range between $1e07-1e11 \, \text{A/m}^2$.
  We resort to electrical pulses of similar character in our simulations as well (see caption to Fig.~\ref{fig_drop_tau_sol}).  

  Another important point to note is that every 
  parameter in the crystal plasticity model does not equally influence the reductions in flow stress. We will highlight those which have the largest  
  influence, as we discuss each mechanism.
  
  \begin{table}[h!]
   \begin{center}
    \footnotesize
    \begin{tabular}{|c|c|} 
      \hline
      Parameter & Value \\
      \hline
      \hline
      $\tau_{sol}$  & $7 \, \text{MPa}$ \\
      \hline
      $d$ & $50 \,\mu\text{m}$ \\
      \hline
      $b_s$ & $ 2.86e-10 \,\text{m} $ \\
      \hline
      $Q_s$ & $1.6e-019 \,\text{J}$ \\
      \hline
      $p,q $ & $1.0$ \\
      \hline
      $v0$ & $1.0e04 \,\text{m/s}$ \\
      \hline
      $i_{slip}$ & $30.0$ \\
      \hline
      $C_{anni}$ & $19.0$  \\
      \hline
      $D_0$  & $1.76e-05 \, \text{m}^2\text{/s}$ \\
      \hline
      $Q_{c}$ & $1.55e-019 \, \text{J}$ \\
      \hline 
    \end{tabular}
    \caption{Values of parameters related to dislocation glide and climb}
     \label{tab_param_dislo}
  \end{center}
   
\end{table}

\begin{table}[h!]
   \begin{center}
   \footnotesize
    \begin{tabular}{|c|c|} 
      \hline
      Parameter & Value \\
      \hline
      \hline
      $j_0$ & $3.0e09 \, \text{A/m}^2$ \\
      \hline
      $\beta$ & $1e-03$ \\
      \hline
      $(\rho_D/N_D)$ & $ 3.3e-025 \, \Omega \text{m}^3 $ \\
      \hline
      $e$ & $1.6e-019 \, \text{C}$ \\
      \hline
      $n_e  $ & $ 1.8e29 \, \text{m}^{-3}$ \\
      \hline
    \end{tabular}
    \caption{Values of parameters related to EP} 
     \label{tab_param_ep}
  \end{center}
  
\end{table}

\begin{table}[h!]
   \begin{center}
    \footnotesize
    \begin{tabular}{|c|c|} 
      \hline
      Parameter & Value \\
      \hline
      \hline
      $\rho$ & $2700 \, \text{kg/m}^3$ \\
      \hline
      $C_p$ & $900 \, \text{J/(kg K)}$ \\
      \hline
      $K$ & $ 204 \,  \text{Wm}^{-1}\text{K}^{-1} $ \\
      \hline
      $\sigma_{el}$ & $3.5e7 \,\text{S/m} $ \\
      \hline
    \end{tabular}
    \caption{Values of electrical and mechanical properties}
     \label{tab_param_el_mech_prop}
  \end{center}
  
\end{table}

 The results presented in this section are in terms of that component of the applied 
 stress tensor which denotes a normal stress along the axis of compression ($\sigma_{appl}$).
 Similarly, the normal component of the imposed strain tensor along the axis of the compression test 
 is referred to as strain ($\epsilon$) in this section.

 In the discussions that follow, we consider each of the softening mechanisms in isolation.
 Such an approach should help to delineate the relative contributions of each of the mechanisms towards the 
 electroplastic effect. As there are several possible sources of softening to be considered when 
 de-pinning of dislocations is the operative mechanism, we follow an order which is identical 
 to that used in Section~\ref{dep_dislo} while discussing them. 
 
 We begin with the case where a change in $\tau_{sol}$ 
 due to de-pinning of dislocations is the operative mechanism causing flow softening.
 Referring to the theoretical discussion in Section~\ref{dep_dislo}, we can see that 
 $j_0$ is a parameter which acts as a normalizing factor to the current density $j$ 
 and hence exerts an influence on the reductions in flow stress obtainable due to dislocation de-pinning.
 In the absence of a suitable experimental dataset to determine $j_0$ by fitting, we will choose
 for its value a quantity which is very similar to that reported for Aluminium~\cite{Molotskii1995, Molotskii2000}.
 The flow curve presented in Fig.~\ref{flow_curve_tau_sol} 
 displays a stress drop of around 
 $3 \, \text{MPa}$ observed coincident with the electrical pulse. The drop is about 
 $12 \%$ of the computed flow stress just prior to the point of application of the pulse. 
 The details of the loading and the pulse character are mentioned in 
 the caption to Fig.~\ref{fig_drop_tau_sol}. 
 
 The crystal plasticity
 extensions to model the softening mechanisms which stem from de-pinning of dislocations
 involve the ratio $(j/j_0)$ as a crucial parameter. 
 In order to explore the behaviour of the reductions in flow stress as a function of the ratio $(j/j_0)$, 
 we record the reductions in flow stress from simulations conducted over a range of  
 values of $(j/j_0)$ and present them in Fig.~\ref{str_drp_tau_sol_func_j}.
 The curve from simulations display several features of Eq.~\ref{str_drop_mol_fin}. These include
 the parabolic nature of the curve and saturation of the reductions in flow stress at small and high values of $(j/j_0)$, respectively. 
 Thus, it can be claimed that the range of $(j/j_0)$ chosen for 
 our analysis is large enough to capture all the key features of the reductions in flow stress. 
 But even as the softening behaviour from simulations 
 qualitatively agrees with the analytical prediction of Eq.~\ref{str_drop_mol_fin}, quantitatively 
 there are differences. The deviation of the simulation curve 
 from that due to Eq.~\ref{str_drop_mol_fin}
 becomes evident when we attempt to fit the simulation data to an expression of the 
 form $f(j/j_0) = A (j/j_0)^2/(1 + (j/j_0)^2)$, with $A$ as the fitting parameter,
 following Eq.~\ref{str_drop_mol_fin}. 
 An important difference between the simulation and the fitted curves is that 
 the point of inflection in the simulation curve 
 no longer manifests at $j/j_0 = 0.577$, but is rather observed at around $0.8$.
 This discrepancy between theory and simulations is not surprising because of two reasons. 
 The first being that of the validity of the scheme which is invoked to derive Eq.~\ref{str_drop_mol_fin}
 where the behaviour of a single grain normalized by a Taylor factor is considered to be representative 
 of the bulk polycrystalline sample. We have already discussed this point in the paragraph following Eq.~\ref{str_drop_mol_fin}. 
 The second reason for a possible discrepancy between theory and simulations is due to the 
 assumed constancy of the average long range elastic stress field ($\sigma_{pass}$) before and during pulsing,
 which is used to derive Eq.~\ref{str_drop_mol_fin} from Eq.~\ref{str_drop_mol_fin_eff}. 
 There is no way to verify this assumption 
 as the average long range stress field ($\sigma_{pass}$) cannot be explicitly 
 written as functions of $\tau_{pass}^\alpha$ which prevents it from being 
 determined computationally and can only be related approximately to $\tau_{pass}^\alpha$
 using the Taylor factor ($M$).   
 
 From Eq.~\ref{str_drop_mol_fin}, a given value of $\tau_{sol}$ 
 is expected to have a large
 bearing on the $\%$ drop in stress. 
 In order to explore the effect of $\tau_{sol}$ further, 
 we focus our attention on the Orowan equation 
 in Eq.~\ref{Orowan_no_pulsing_bulk} and notice that a particular value 
 of externally imposed $\dot{\epsilon}$ 
 is satisfied by a certain ratio of $(\sigma_{eff}/\tau_{sol})$. 
 Thus, when $\tau_{sol}$ changes $\sigma_{eff}$ should also change to maintain 
 the ratio ($\sigma_{eff}/\tau_{sol}$) constant, implying 
 a direct proportionality between $\sigma_{eff}$ and $\tau_{sol}$.
 Using this fact along with a direct proportionality 
 between  $\Delta \sigma_{appl}$ and
 $\sigma_{eff}$ from Eq.~\ref{str_drop_mol_fin} 
 translates into a similar
 scaling between $\Delta \sigma_{appl}$
 and $\tau_{sol}$. Thus, the reductions in flow stress $\Delta \sigma_{appl}$
 should scale linearly with $\tau_{sol}$ and this could indeed be confirmed from 
 Fig.~\ref{str_drp_tau_sol_func_tau_sol} which shows a straight line 
 relationship to exist between $\Delta \sigma_{appl}$ computed from simulations 
 for different values of $\tau_{sol}$ but at a particular value of $j$.
 We have also fitted the simulation data with a straight line and found the slope to be $0.44$.
 In view of the importance of $\tau_{sol}$  in determining the reductions in flow stress, 
 we present a short discussion 
 in the Appendix describing the rational behind the selection of a suitable 
 value for our simulations.

\begin{figure}[!htbp]
  \begin{center}
   \subfigure[]{\includegraphics[width=0.8\linewidth]{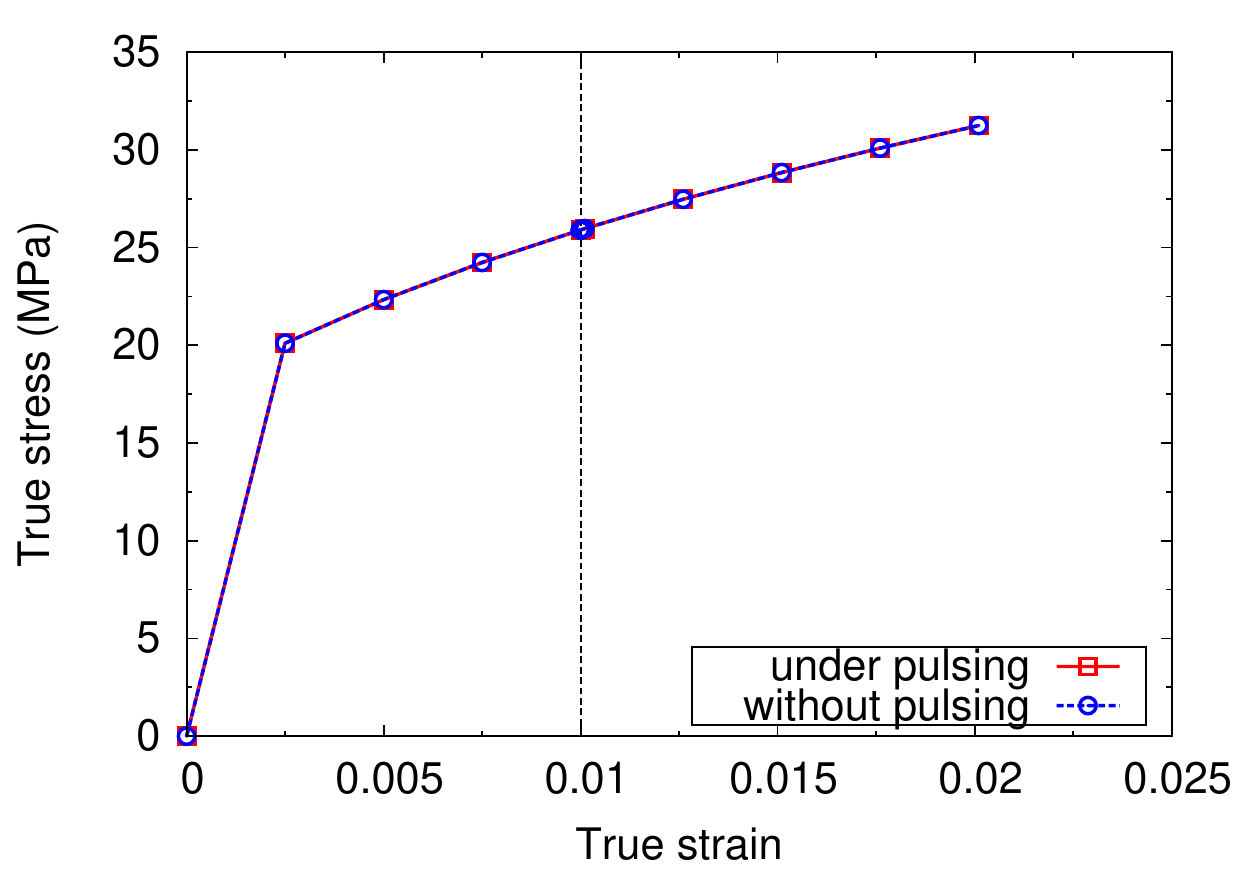}
    \label{flow_curve_lambda_slip_drop}}
   \subfigure[]{\includegraphics[width=0.8\linewidth]{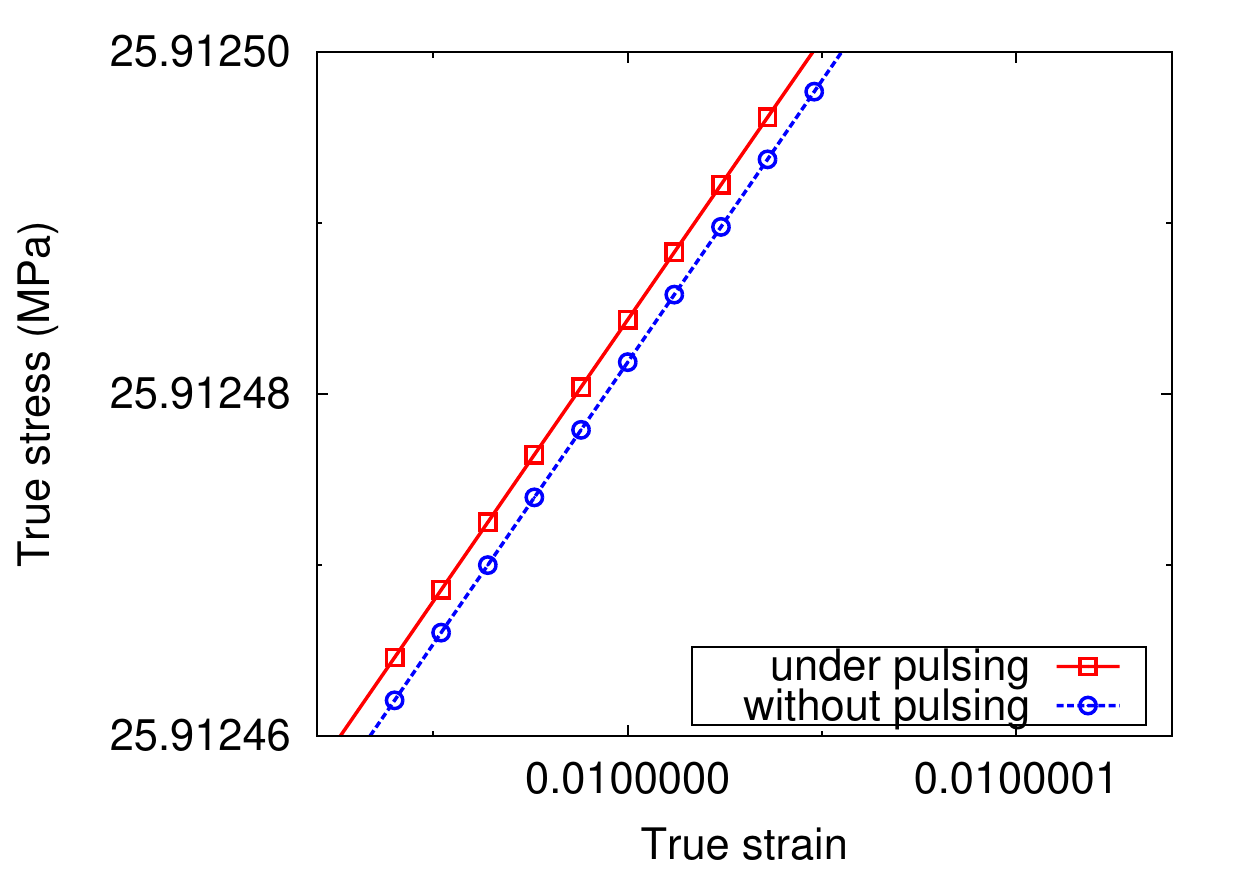}
    \label{flow_curve_lambda_slip_drop_mag}} 
  \end{center}
  \caption{Figures demonstrating reductions in flow stress due to a change in $\lambda_{slip}^\alpha$. 
  (a) Flow curves with and without electropulsing during a compression test.
  (b) The magnified version of (a) around a true strain of $0.01$.
   The loading and pulsing details are mentioned in the caption to Fig.~\ref{fig_drop_tau_sol}.}
  \label{fig_drop_lambda_slip}
 \end{figure}

 Moving on to the second possible source of reductions in flow stress which is an increase in $\lambda_{slip}^\alpha$,
 the corresponding flow curve in Fig.~\ref{flow_curve_lambda_slip_drop} hardly shows a difference from the 
 un-pulsed one. A slight rise in the flow stress of $\approx 1e-06 \, \text{MPa}$ can be discerned when we magnify the flow curves around the strain
 at which the pulse has been applied as seen in Fig.~\ref{flow_curve_lambda_slip_drop_mag}. 
 A rise in flow stress during electropulsing is consistently observed for this particular mechanism
 across the entire range of $(j/j_0)$ considered in Fig.~\ref{str_drp_tau_sol_func_j}.
 But as the increase in stresses are in the same range as the errors due to 
 numerical discretization and precision, a clear trend does not 
 emerge in the variation of $\Delta \sigma_{appl}$ versus $(j/j_0)$, 
 like observed in Fig.~\ref{str_drp_tau_sol_func_j}.
 The possibility of an increase of the flow stress instead of a drop 
 when $\lambda_{slip}^\alpha$ increases has been
 discussed in Section~\ref{sec_lambda_slip}.

\begin{figure}[!htbp]
  \begin{center}
   \includegraphics[width=0.8\linewidth]{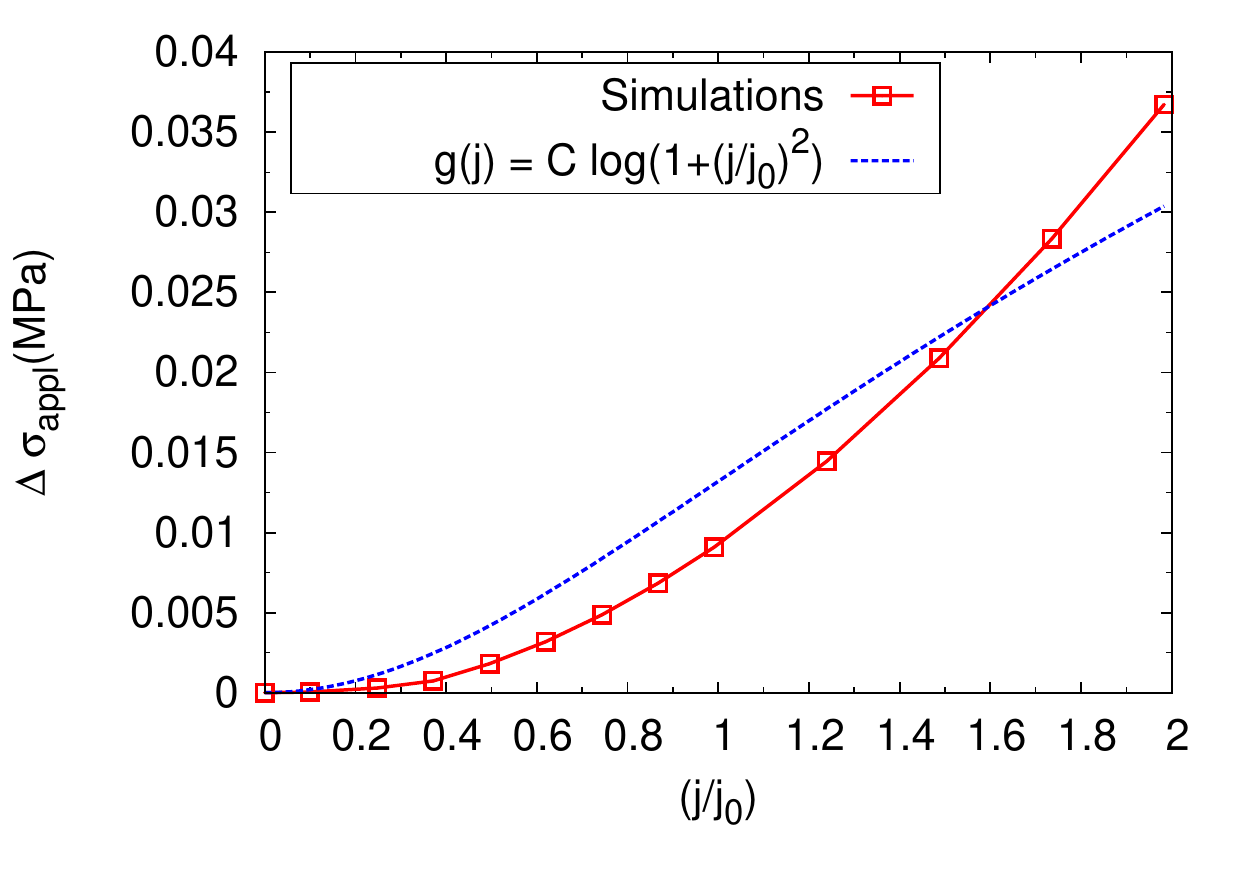}
  \end{center}
  \caption{Figure demonstrating reductions in flow stress due to a change in $v_0$ as a function of $(j/j_0)$.
   The loading and pulsing details are mentioned in the caption to Fig.~\ref{fig_drop_tau_sol}. The 
   constant $C$ is determined by fitting to the simulation data.}
  \label{fig_drop_v0}
 \end{figure}

 We now consider the effects of a change in $v_0$ on the flow curve. 
 A flow curve for this situation is no longer presented 
 as it resembles Fig.~\ref{flow_curve_lambda_slip_drop} and instead we just present a variation 
 of the reductions in flow stress as a function of $(j/j_0)$ in Fig.~\ref{fig_drop_v0}. 
 It is clear that the reductions in flow stress due to a change in $v_0$ are about two orders of magnitude lower 
 than those observed for the case where a change in $\tau_{sol}$ is the 
 source of softening. Also, the reductions in flow stress appear to be a parabolic function of $(j/j_0)$ and do 
 not display a good fit with an expression of the form predicted by Eq.~\ref{str_drop_v0}. The reasons 
 for this deviation are the same as the ones mentioned during the discussion of the reductions in flow stress due to 
 a change in $\tau_{sol}$. 
 
\begin{figure}[!htbp]
 \begin{center}
  \includegraphics[width=0.8\linewidth]{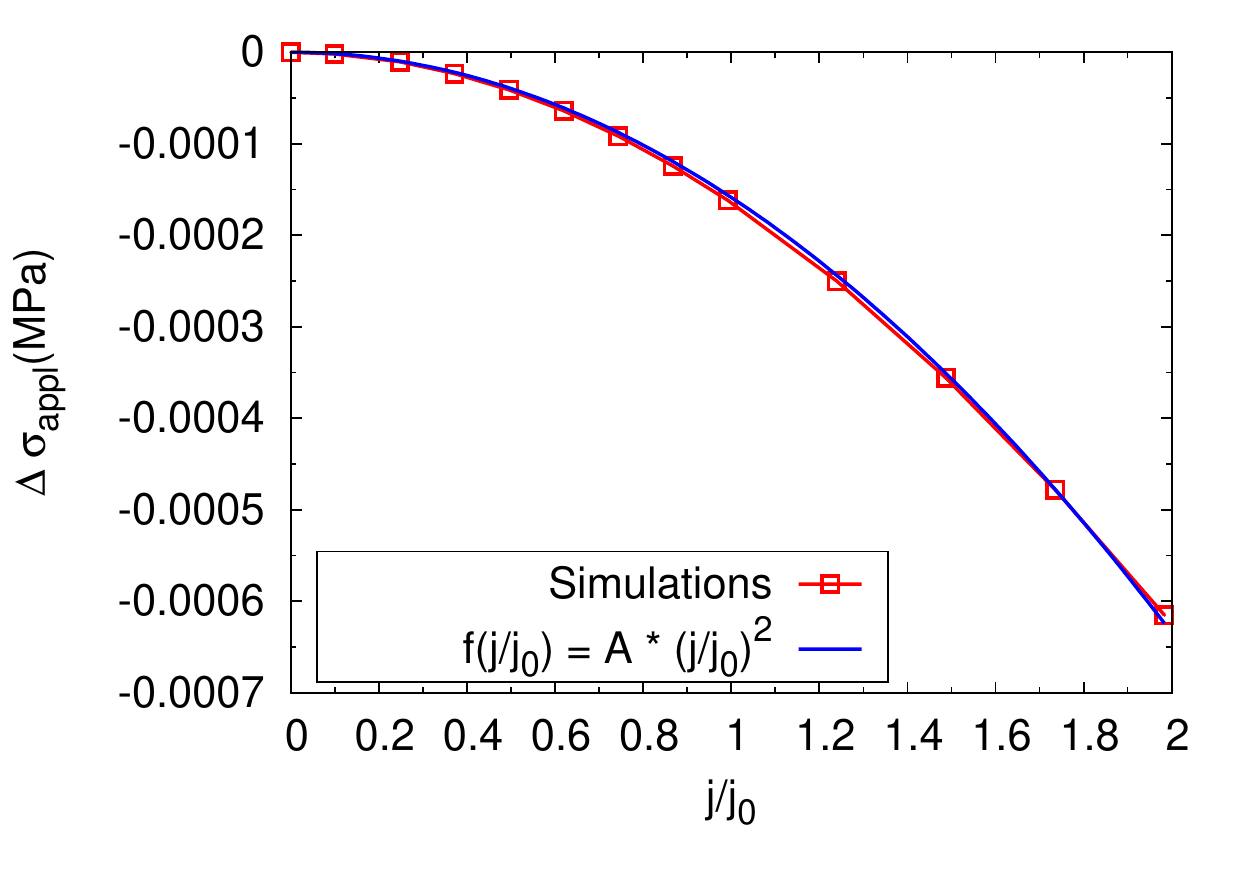}
 \end{center}
 \caption{Figure demonstrating stress rises due to a change in $\tau_{pass}^\alpha$ as a function of $(j/j_0)$.
 The loading and pulsing details are mentioned in the caption to Fig.~\ref{fig_drop_tau_sol}.}
  \label{str_drp_tau_pass_func_j}
 \end{figure}

 This brings us to the final source of change
 when dislocations are de-pinned from obstacles,
 reflected by a change in $\tau_{pass}^\alpha$.
 The value of the parameter $\beta$ in Eq.~\ref{str_drop_tau_pass} is chosen 
 to be of the same order of magnitude as that reported in~\cite{Molotskii1996}(see Table.~\ref{tab_param_ep}).
 As discussed earlier, when de-pinned, larger free
 length of dislocations respond strongly to the 
 elastic stress fields due to other dislocations which lead to larger strain hardening. Thus, 
 this is a mechanism which always leads to an increase in flow stress during pulsing as
 confirmed from Fig.~\ref{str_drp_tau_pass_func_j} where the reductions in flow stress are 
 presented as functions of $(j/j_0)$. The nature of the curve is parabolic and 
 hence allows a close fit by a function $f(j/j_0)= A (j/j_0)^2$ following Eq.~\ref{str_drop_tau_pass}
 where $A$ is a fitting constant.
 The close fit between the form of Eq.~\ref{str_drop_tau_pass} and those observed in Fig.~\ref{str_drp_tau_pass_func_j}
 is rather fortuitous given the assumptions of the analytical predictions.  
 The maximum value of the rise in $\Delta\sigma_{appl}$ over the range of current densities considered 
 is insignificant compared to a drop caused by a change in $\tau_{sol}$.
 
\begin{figure}[!htbp]
 \begin{center}
  \includegraphics[width=0.8\linewidth]{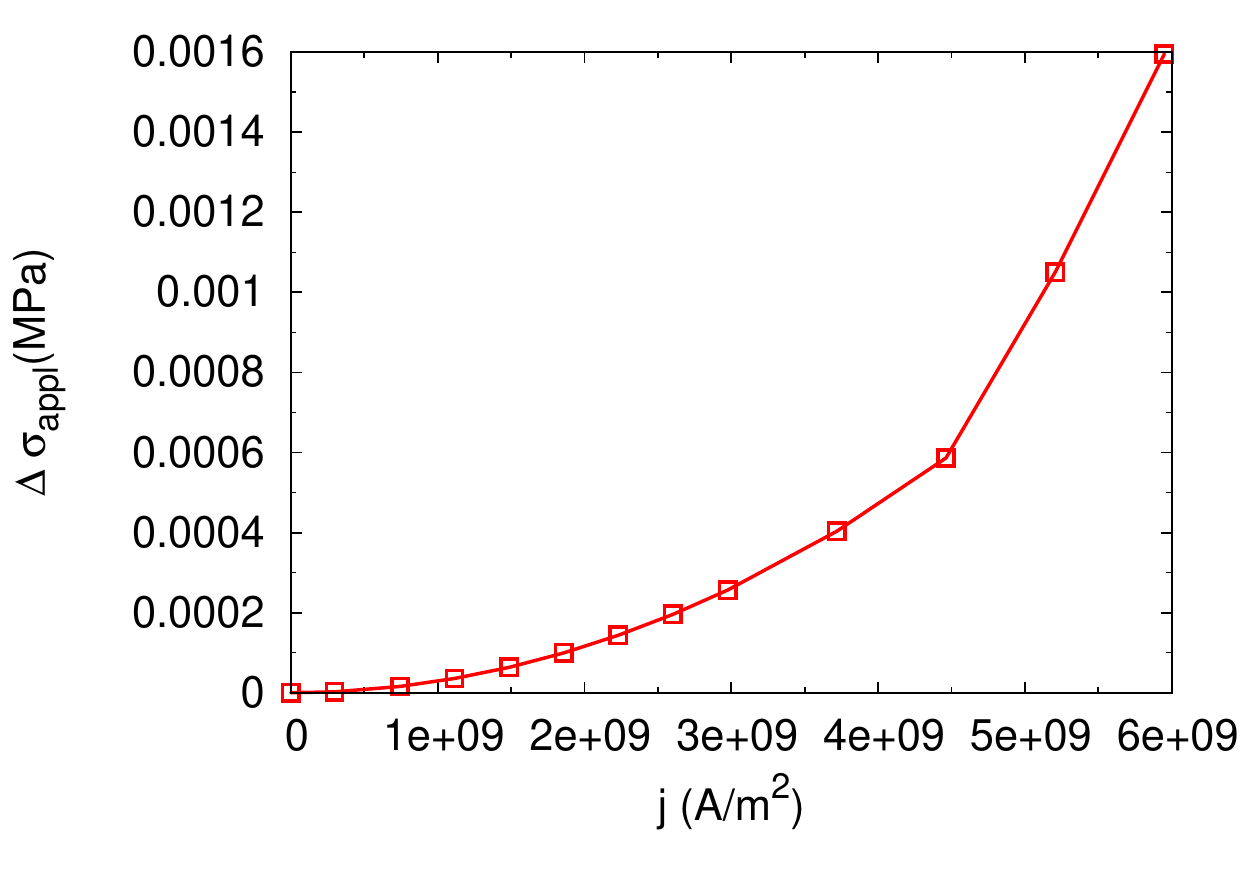}
 \end{center}
 \caption{Figure demonstrating the reductions in flow stress due to electron-wind force as a function of $j$. 
 The loading and pulsing details are mentioned in the caption to Fig.~\ref{fig_drop_tau_sol}.}
   \label{str_drp_el_wind_j}
 \end{figure}

 After dealing with the softening sources which are due to the de-pinning of dislocations 
 from obstacles during electropulsing, we now focus on the effects of electron-wind force 
 on dislocations. The variation of the reductions in flow stress due to electron-wind force
 $\Delta \sigma_{appl}$ as a function of $j$
 has a parabolic character initially but displays a sharper 
 change at higher values of $j$ as evident from the last three points of the curve 
 presented in Fig.~\ref{str_drp_el_wind_j}. Such a behaviour correlates well with the 
 nature of the '$\cosh$' function which is introduced as a factor in Eq.~\ref{Orowan_eq_elec_wind_final}.
 As claimed by Molotskii et al.,~\cite{Molotskii1995},
 we see a very small stress drop of the order of $1e-03 \, \text{MPa}$ due to this mechanism.
 The selection of the relevant parameters like $\rho_D/N_D$ and $n_e$ which control
 the softening through this particular mechanism is described in the Appendix and their values are presented 
 in Table~\ref{tab_param_ep}.
  
\begin{figure}[!htbp]
 \begin{center}
  \subfigure[]{\includegraphics[width=0.8\linewidth]{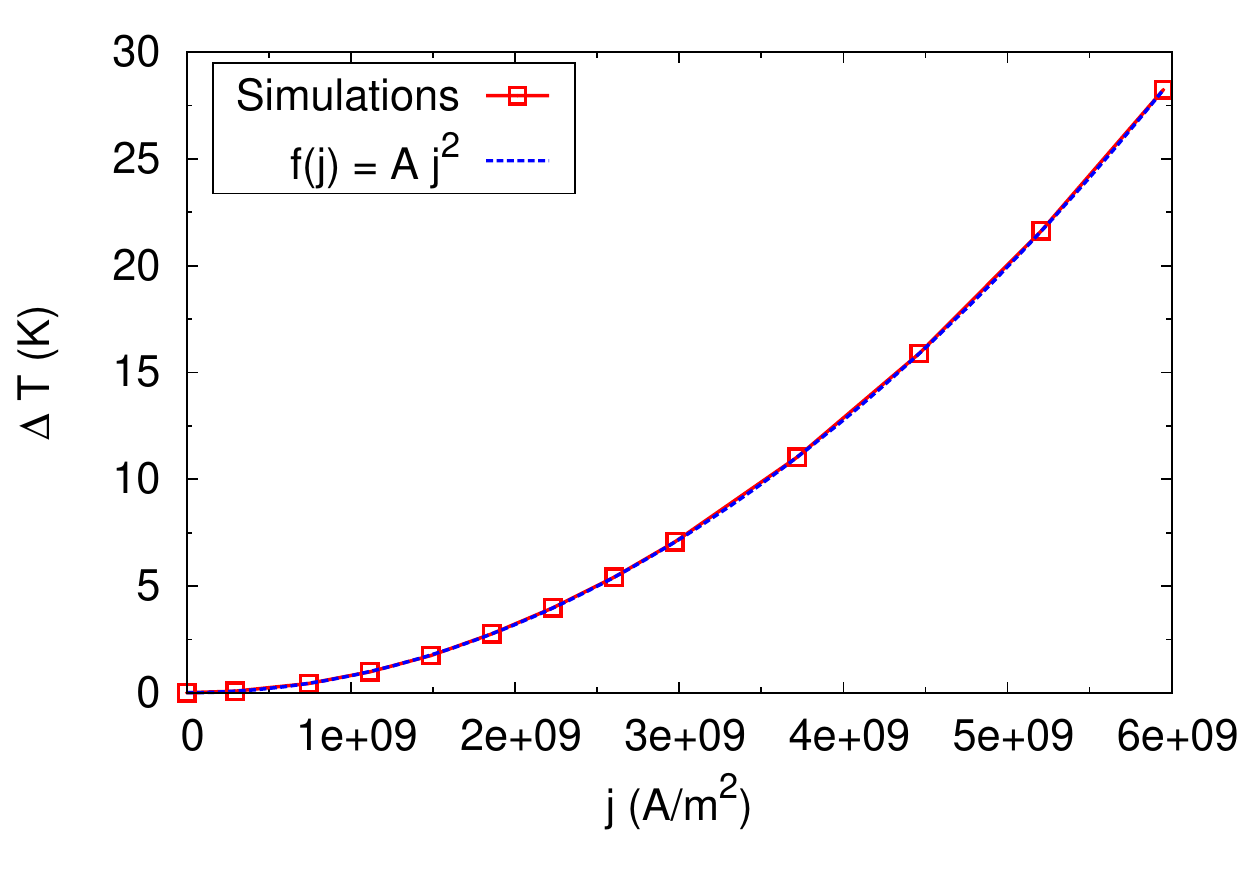}
   \label{temp_rise_jh_j}}  
  \subfigure[]{\includegraphics[width=0.8\linewidth]{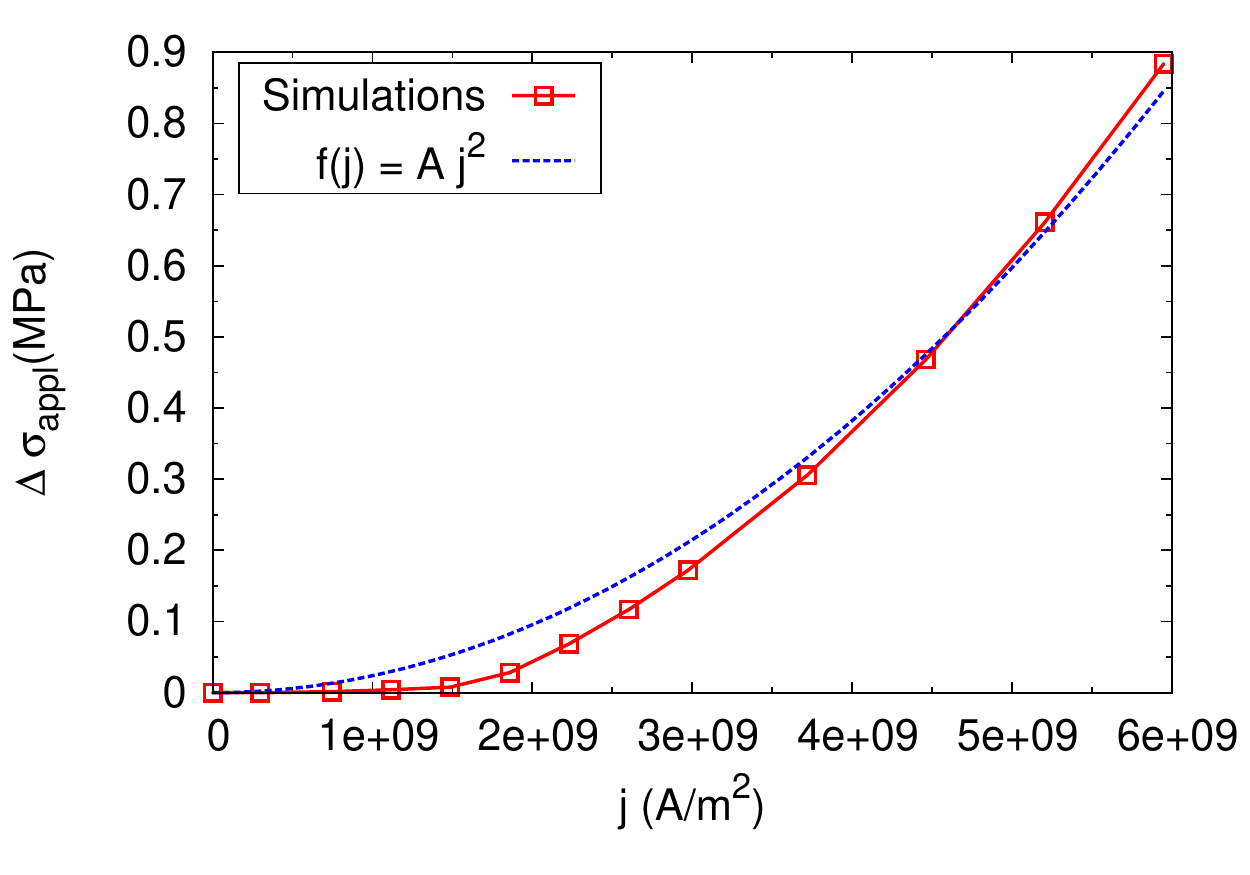}
   \label{str_drp_jh_j}} 
 \end{center}
 \caption{Figures demonstrating the reductions in flow stress due to Joule-heating. 
 (a) Temperature rise as a function of $j$.
 (b) Reductions in flow stress as a function of $j$. 
 The parameter $A$ in the figure legends denote constants which have to be determined 
 by fitting to simulation data.
 The loading and pulsing details are mentioned in the caption to Fig.~\ref{fig_drop_tau_sol}.}
 \label{fig_drop_jh}
 \end{figure}

 Through the results presented so far, 
 we have developed an understanding of the nature and 
 magnitude of the reductions in flow stress due to the 
 de-pinning of dislocations and the electron-wind force. 
 These mechanisms induce some changes either in the 
 interaction between dislocations and obstacles
 or modify the forces acting on dislocations. 
 In other words, these mechanisms  do not 
 invoke a change in temperature of the material
 and hence are athermal in character.
 It is important to determine the quantum of the reduction in flow stress achievable
 due to Joule heating of the material and compare it against 
 the softening observed from athermal means.
 The first result in this regard is a variation of  
 temperature rise ($\Delta T$) with $j$ as presented in Fig.~\ref{temp_rise_jh_j}.
 The corresponding reductions in flow stress are presented in Fig.~\ref{str_drp_jh_j}.
 The variation of temperature rise with $j$ follows a parabolic curve which is in accordance 
 with our analysis expressed by Eq.~\ref{temp_sim}.
 The nature of the curve in Fig.~\ref{str_drp_jh_j}
 is parabolic for $j> 2e09 \, \text{A/m}^2$ and is in 
 accordance with that predicted by Eq.~\ref{str_drp_jh_j}.
 For $j<2e09 \, \text{A/m}^2$, the softening response seen in simulations is lower than 
 that predicted by our analysis. Such values of $j$ correspond to low temperature
 rises of $\Delta T < 5\text{K}$, which can be conjectured to be not large enough 
 to cause significant lowering of flow stress due to enhanced thermal activation.
 In other words, it appears, that unless the temperature exceeds a certain threshold value, 
 there is no significant softening due to Joule-heating.
 This brings the limitation of analytical expressions like Eq.~\ref{str_drp_jh_j} again
 to the forefront as the assumptions involved in deriving such expressions
 are too simplistic compared to actual crystal plasticity simulations. The reductions in flow stress
 $\Delta \sigma_{appl}$ observed due to Joule-heating are higher than all the other softening 
 sources discussed above except for the case where a change in $\tau_{sol}$ 
 causes softening. It can be noted that the activation energies for slip ($Q_s$) 
 and climb ($Q_c$) are the key parameters which control the reductions in flow stress due to Joule-heating.
 We have described the process of selection of $Q_s$ in the Appendix, and from Fig.~\ref{fig_drop_jh}
 it is reasonable to argue that the rise in temperature is not enough to cause significant climb
 of edge dislocations. 
 
\begin{figure}[!htbp]
 \begin{center}
  \includegraphics[width=0.8\linewidth]{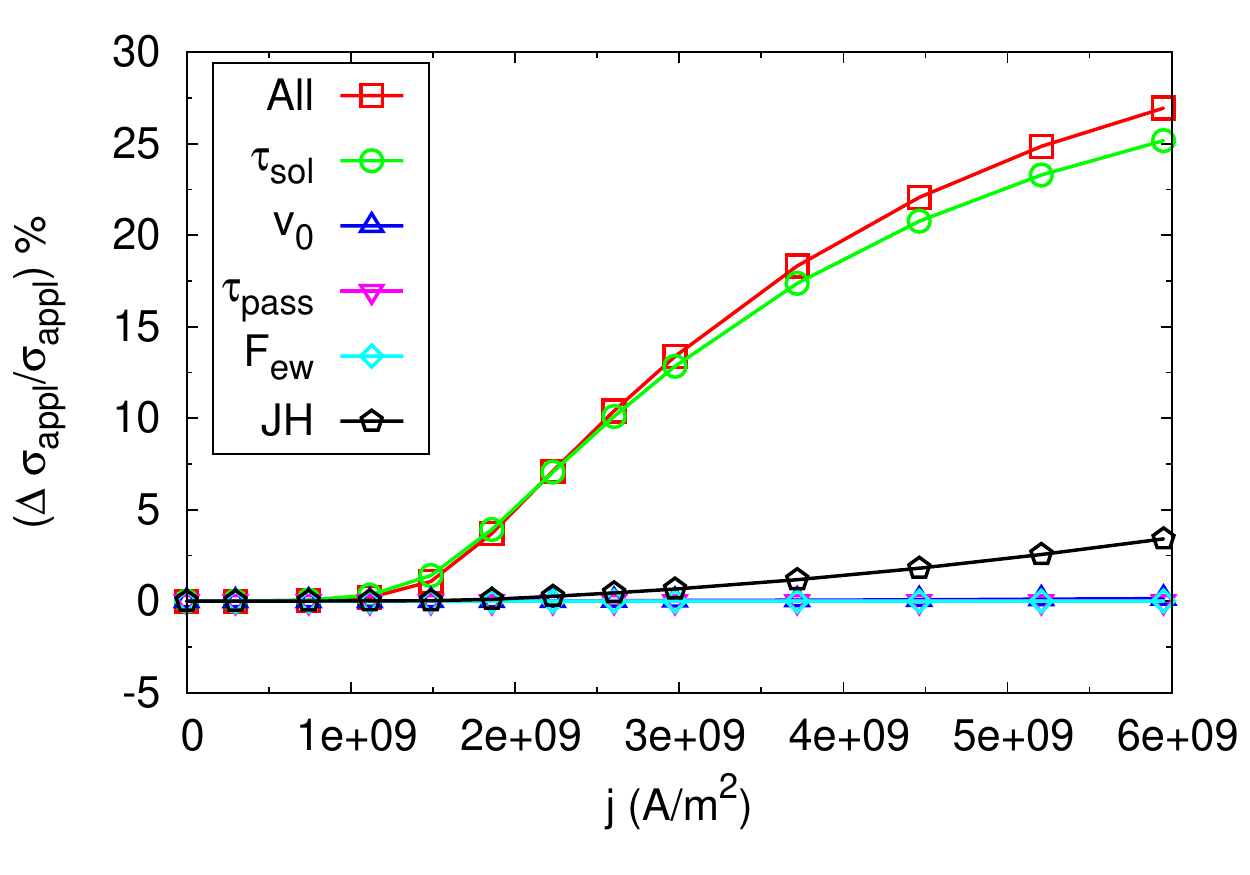}
 \end{center}
 \caption{Figure comparing the $\%$ reductions in flow stress due to all the mechanisms, plotted as a function of $j$.
 In the figure legends, JH denotes Joule-heating, $F_{ew}$ denotes electron-wind force, and 'All' 
 refers to all mechanisms of EP being active. $\tau_{sol}$ and $v_0$ represent reductions in flow stress due to
 changes in those terms.}
 \label{stress_drop_fraction_all}
 \end{figure}

 We have presented and discussed all the mechanisms and their corresponding 
 sources which result in the electroplastic effect. A graphical summary
 of our observations is presented in Fig.~\ref{stress_drop_fraction_all} 
 where the $\%$ drops in stress are plotted as function of the current density $j$.
 These figures unambiguously point to modifications in $\tau_{sol}$ being the strongest contributor to the 
 reductions in flow stress. Thermal softening due to joule heating is a distant second, while all the other mechanisms 
 produce reductions in flow stress which are at least smaller by an order of magnitude compared to Joule-heating. 
 A curve obtained from a simulation 
 where all the effects are simultaneously active (denoted by a legend ``All'' in the 
 figure) lies in close proximity to the $\tau_{sol}$ curve and mimics its shape confirming 
 changes in $\tau_{sol}$ to be  the largest contributor to the reductions in flow stress. 
  
\begin{figure}[!htbp]
 \begin{center}
  \subfigure[]{\includegraphics[width=0.8\linewidth]{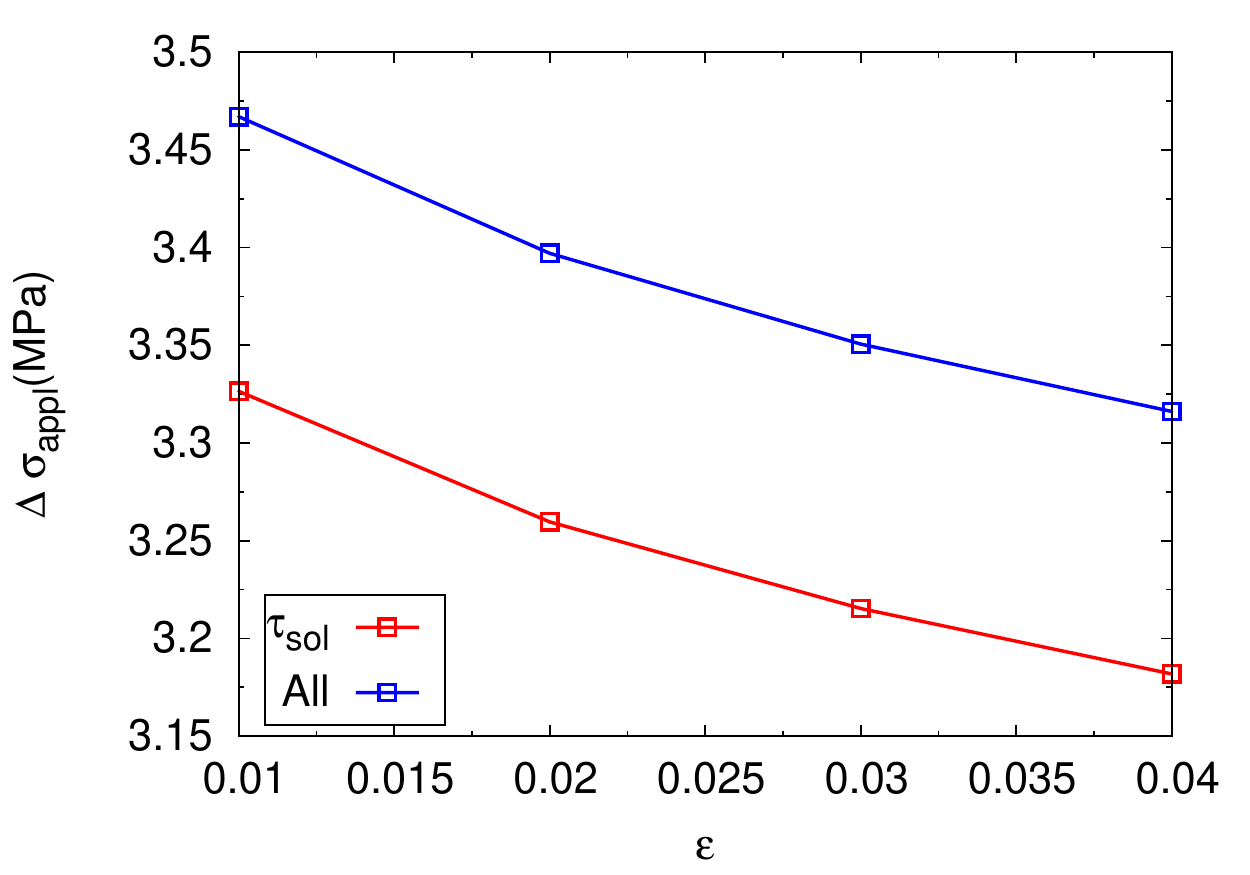}
   \label{tau_sol_drop_with_strain}}
  \subfigure[]{\includegraphics[width=0.8\linewidth]{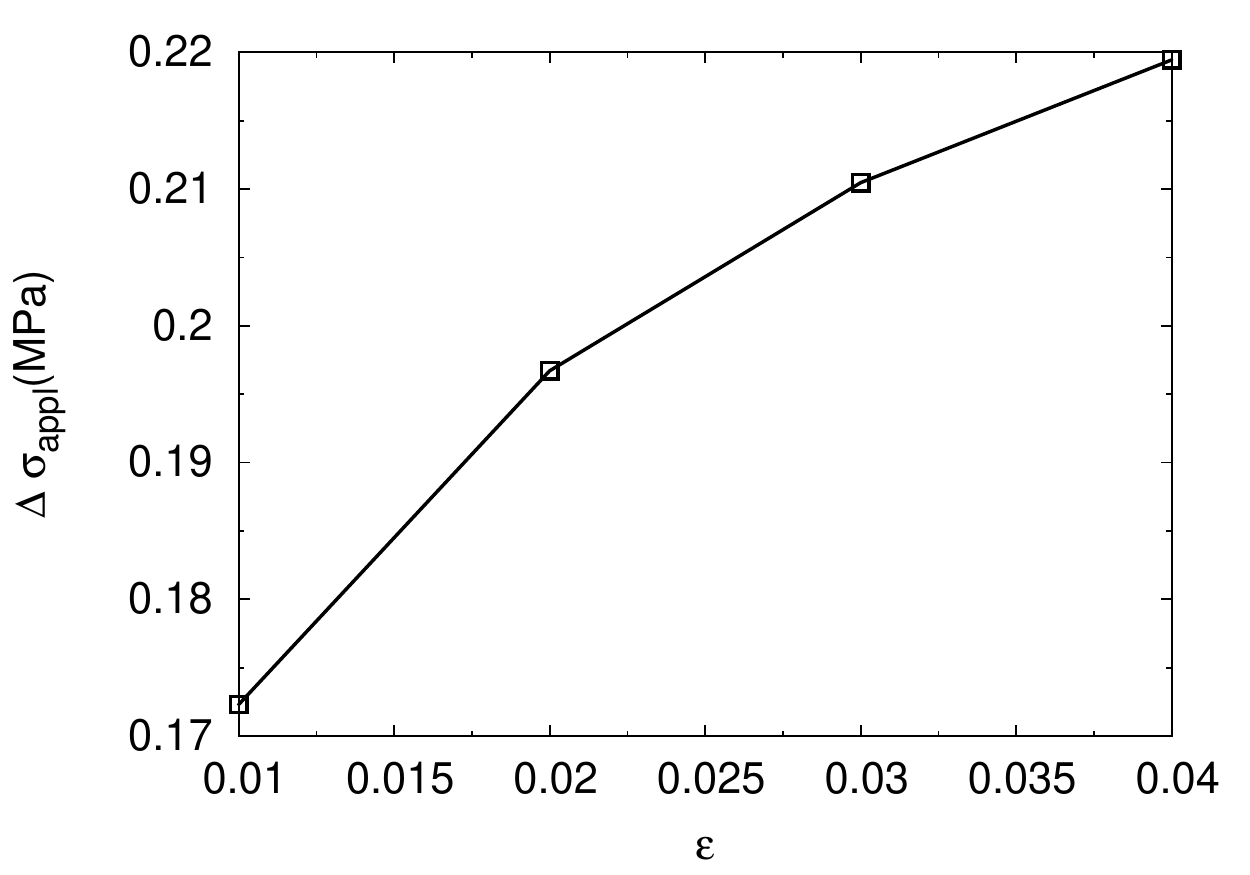}
   \label{JH_drop_with_strain}} 
 \end{center}
 \caption{Figures demonstrating the reductions in flow stress ($\Delta \sigma_{appl}$) due to, (a) a change in $\tau_{sol}$ and when all the mechanisms are active,
 and (b) Joule-heating, with strains ($\epsilon$) at which electrical pulses are applied. The figure legends are explained in the caption 
 to Fig.~\ref{stress_drop_fraction_all}. Pulses of magnitude $3e09 \, \text{A/m}^2$ are applied for a total time of $60 \, \mu\text{s}$.}
 \label{str_drp_with_strain}
 \end{figure}

 There is a considerable experimental evidence that the magnitudes of the reductions in flow stress are reduced
 as the the pulses are applied at higher values of strains~\cite{Sprecher1986, Conrad2002, Molotskii2000}. In Figs.~\ref{tau_sol_drop_with_strain}
 and~\ref{JH_drop_with_strain}, we present a variation of the reductions in flow stress due to a change in $\tau_{sol}$ and Joule-heating
 respectively,  as functions of the strains at which the sample is pulsed.
 It can be seen from Fig.~\ref{tau_sol_drop_with_strain} that $\Delta \sigma_{appl}$ falls 
 as pulses are applied at higher strains. This could be explained in the following manner.
 In our crystal plasticity framework, higher strains are  microstructurally characterized by a larger value of $\rho_m^\alpha$ (and also $\rho_d^\alpha)$.
 Invoking the assumption that an imposed $\dot{\epsilon}$ leads to constant values of $\dot{\gamma^\alpha}$ on the different slip planes
 $\alpha$, from Eq.~\ref{Orowan_eq} it is clear that for larger values of $\rho_m^\alpha$ at higher strains, 
 smaller values of $\tau_{eff}^\alpha$ satisfy Eq.~\ref{Orowan_eq}.
 As $\sigma_{eff}$ is related to $\tau_{eff}^\alpha$ by a Taylor factor,
 the value of $\sigma_{eff}$ will also be lowered as the strains increase. 
 The direct proportionality between the reductions in flow stress ($\Delta \sigma_{appl}$)
 due to a change in $\tau_{sol}$ and $\sigma_{eff}$ due to Eq.~\ref{str_drop_mol_fin}
 explains the lowered reductions in flow stress ($\Delta \sigma_{appl}$) at higher values of strains.
 In other words, at higher levels of strains, plasticity is achieved by generating more 
 mobile dislocations to compensate for the smaller dislocation free paths. 
 Thus, at higher strains, the mechanisms of EP which aid thermal 
 activation of the dislocation segments over short range obstacles can only 
 enhance the much lower dislocation velocity by a smaller factor,
 compared to that possible at smaller strains.
  
 In contrary to our observations in Fig.~\ref{tau_sol_drop_with_strain}, the reductions in flow stress due to 
 thermal softening increase with applied strains as shown in Fig.~\ref{JH_drop_with_strain}.
 Using the concept of the reduction in $\sigma_{eff}$ with strain as described in the previous paragraph,
 the increase in reductions in flow stress observed in Fig.~\ref{JH_drop_with_strain} could be immediately explained
 from Eq.~\ref{str_drp_therm_sof}.
 But as these changes are an order of magnitude smaller 
 than those observed in Fig.~\ref{tau_sol_drop_with_strain},
 the  curve corresponding to the case 
 where all the softening mechanisms are operative, mimics the one
 representing reductions in flow stress due to a change in $\tau_{sol}$ (see Fig.~\ref{tau_sol_drop_with_strain}).
 The influence of the other mechanisms of EP are not considered to be important 
 because they have been seen to produce a negligibly small impact on the reductions in flow stress. 
 
\begin{figure}[!htbp]
 \begin{center}
  \includegraphics[width=0.8\linewidth]{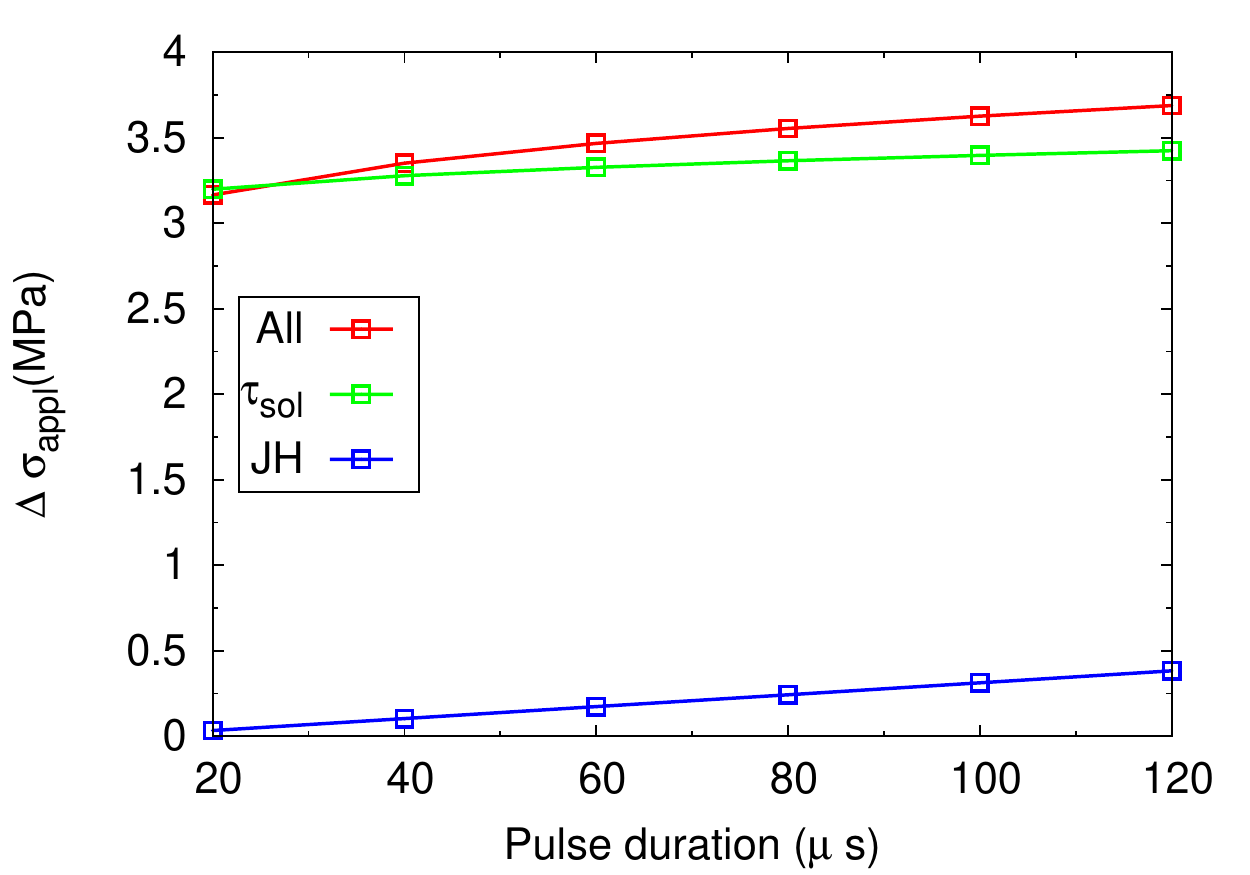}
 \end{center}
 \caption{Figure demonstrating the variation of the reductions in flow stress as a function  
  of the pulse width. The current density is maintained at $j=3e09 \, \text{A/m}^2$. The figure legends are explained in the caption 
 to Fig.~\ref{stress_drop_fraction_all}. }
 \label{stress_drops_with_pulse_width}
 \end{figure}

 Finally, we probe the effect of pulse duration on the reductions in flow stress as presented 
 in Fig.~\ref{stress_drops_with_pulse_width}, where 
 the pulse duration is increased keeping the current density constant. 
 The reductions in flow stress due to a change in $\tau_{sol}$ increase gradually 
 as the pulse durations are increased and show signs of saturation at higher pulse durations. 
 This can be explained by discretizing the entire pulse duration into 
 a series of infinitesimal pulses, each lowering the $\sigma_{eff}$ to a value
 which becomes the initial $\sigma_{eff}$ for the subsequent pulse (see Eq.~\ref{str_drop_mol_fin_eff}).
 Noting that $\sigma_{eff} << \sigma_{appl}$ and $(\sigma_{eff}/M)<\tau_{sol}$,
 the proportionality of $\Delta \sigma_{appl}$ with $\sigma_{eff}$ (see Eq.~\ref{str_drop_mol_fin})
 means that the reductions in flow stress increase with the pulse size, but ultimately saturates.
 Thermal softening due to Joule heating is strongly affected
 as more electrical energy is introduced into the material as the 
 pulse duration increases. Joule heating being the mechanism which is most
 sensitive to an increase in pulse duration dominates 
 the overall sensitivity of the electroplastic effect even when all the mechanisms 
 are active as seen in Fig.~\ref{stress_drops_with_pulse_width}.
 
 Having reached the end of this section, we will summarize our key observations:
 \begin{itemize}
  \item De-pinning of dislocations produces the largest reductions in flow stress through a change in $\tau_{sol}$. 
  \item Joule-heating is the second largest contributor to the reductions in flow stress but its contribution 
        is still an order of magnitude smaller than that produced by dislocation de-pinning. 
  \item Electron-wind force has negligible contribution to the reductions in flow stress.
  \item The reductions in flow stress due to de-pinning of dislocations fall as the electrical pulses are applied at higher strains.
  \item Increasing duration of the electrical pulses increases the reductions in flow stress due to both Joule-heating and dislocation de-pinning.
  \item As the reductions in flow stress due to de-pinning of dislocations are a strong function of $\tau_{sol}$, 
        it can be expected that for certain combinations of parameters, 
        Joule-heating produces larger reductions in flow stress than that due to de-pinning of dislocations. 
 \end{itemize} 
  
 In the next section, we discuss a few implications of our results and lay out 
 the possibilities for future work in this direction.
   
 \section{Discussion}
 As already discussed, for the parameter set considered, 
 dislocation de-pinning produces a reduction in flow stress which is an order 
 of magnitude larger than that produced by Joule-heating. 
 But our simulations also demonstrate that it is possible for 
 Joule-heating to supersede the de-pinning of dislocations as the dominant mechanism of EP.
 This happens when the temperature rise produced in the material is large enough to cause a softening 
 higher than the softening produced by de-pinning of dislocations.
 In context of this understanding, a few experimental evidences which 
 claim EP to be predominantly a Joule-heating phenomenon can 
 now for the first time be explained consistently within the framework of our model.
 Examples of such studies are due to 
 Magargee et al.,~\cite{Magargee2013} and Zheng et al.,~\cite{Zheng2014}.
 It must be noted that the material under consideration in these studies is Ti,
 which has a hexagonal close packed structure. The fact that 
 there are not enough slip systems in a hexagonal structure 
 to produce a predominantly dislocation mediated plasticity 
 also implies that dislocations interact with forest 
 dislocations much rarely in such materials resulting in 
 smaller values of $\tau_{sol}$, compared to those observed for FCC materials.
 Hence, it is possible that for such materials,
 de-pinning of dislocations during electropulsing may not be 
 the dominant softening mechanism. 
 This point has also been raised by Sprecher et al.,~\cite{Sprecher1986} where they claim that 
 the interstitial impurities in Ti present the largest barriers to dislocation motion for hexagonal metals. 
 So, for such a material it is quite reasonable that thermal softening 
 due to Joule-heating is the dominant mechanism for EP, especially 
 when the interstitial solutes are not paramagnetic in character.  
 Another work by Goldman et al.,~\cite{Goldman1981} reports no softening
 in Pb at superconducting temperatures (around $4K$). 
 This could be explained by considering that a change in electronic behaviour 
 which induces superconductivity at low temperatures could also potentially
 inhibit the mechanism of de-pinning of dislocations, which is dependent on 
 electronic states. 
   
 With reference to the discussion in the previous paragraph,
 it can be noted that 
 while the mechanism of de-pinning 
 of dislocations is of central importance to the phenomenon of EP 
 in FCC materials and may be of significance for Hexagonal materials as well, 
 this is not valid for BCC materials, as has already been pointed out by Molotskii~\cite{Molotskii2000}. 
 For BCC materials, slip is largely determined by the well known kink-pair mechanism, 
 and the way that mechanism is affected by electropulsing continues to remain unclear.
 Hence some experimental suggestions are crucial
 to formulate an athermal theory of EP for BCC materials.

 \section{Acknowledgement}
 The authors gratefully acknowledge the financial support by the DFG Priority Programme SPP 1959/1, 
 "Manipulation of matter controlled by electric and magnetic field: Toward novel synthesis 
 and processing routes of inorganic materials” with the grant number of 319419837.
 
 \section{Appendix}
Here, we present a short description of the
process of selection of different parameters 
which influence the reductions in flow stress during electropulsing.
We first describe selection of $\tau_{sol}$ 
which is central to the reductions in flow stress
observed due to dislocation de-pinning. 
The parameters critical to electron-wind 
force $\rho_D/n_d$ and $n_e$ are discussed next.

\subsection{Selection of $Q_s$ and $\tau_{sol}$}
The activation energy for slip $Q_s$ is related to the energy of forming a jog
as a dislocation intersects a forest dislocation. This energy has been estimated to be
between $Gb_s^3/5$ and $Gb_s^3/3$~\cite{Hull2001} and we assume it to be,
\begin{align}
 Q_s = \dfrac{1}{4} Gb_s^3.
 \label{Qs_estimate}
\end{align}
For fcc materials, it is often seen that $G b_s^3\approx 4\text{eV}$~\cite{Hull2001},
hence $Q_s = 1eV= 1.6e{-}19\text{J}$ from Eq.~\ref{Qs_estimate}. 
When a jog is created by an applied stress $\tau^*$ we can write,
\begin{align}
 Q_s = \tau^* V,
 \label{str_jog_cr}
\end{align}
where $V$ is the activation volume defined as,
\begin{align}
 V = b_s d_{ac} l_c,
 \label{def_activ_volume}
\end{align}
with $d_{ac}$ and $l_c$ denote the activation distance and average free dislocation segment length,
respectively. For jog formation during forest dislocation interaction, $d_{ac} \approx b_s$~\cite{Hull2001},  
and $l_c \approx 1000 b_s$~\cite{Evans1969}. Assuming $b_s=2.86e{-}10 \,\text{m}$, using Eqs.~\ref{str_jog_cr} and~\ref{def_activ_volume}, 
we get, $V\approx 1000 b_s^3$ and $\tau^* \approx 7\,\text{MPa}$. Conceptually, $\tau^*$ is 
the same as $\tau_{sol}$ as the latter represents the stress required for dislocations to overcome 
short range obstacles like forest dislocations. This exercise allows us to identify two important
parameters for our crystal plasticity simulations: $\tau_{sol} = 7\,\text{MPa}$ and $Q_s = 1.6e{-}19\,\text{J}$. 

\subsection{The parameters influencing the electron-wind force}
The two key parameters which affect the electron-wind force are the specific dislocation resistivity $\rho_D/N_D$
and the free electron density $n_e$.  For the former, we use the value stated for Aluminium in ~\cite{Basinski1963}
while we compute $n_e$ using the formula,
\begin{align}
 n_e = \dfrac{f N_A \rho}{M_A},
 \label{free_elec_density}
\end{align}
where, $M_A$ is the atomic mass, $f$ is the number of free electrons per atom, 
$N_A$ is the Avogadro number and $\rho$ is the density. 
We compute $n_e$ using the parameters for Aluminium.

\bibliography{ep_paper_ref.bib}

\end{document}